\documentclass[twocolumn]{aastex63}

\usepackage[T1]{fontenc}

\shorttitle{Coldest Backyard Worlds Brown Dwarfs}
\shortauthors{Meisner et al.}

\graphicspath{{./}{}}

\begin{document}

\title{\textit{Spitzer} Follow-up of Extremely Cold Brown Dwarfs Discovered by the Backyard Worlds: Planet 9 \\ Citizen Science Project}

\correspondingauthor{Aaron M. Meisner}
\email{ameisner@noao.edu}

\author[0000-0002-1125-7384]{Aaron M. Meisner}
\affiliation{NSF's National Optical-Infrared Astronomy Research Laboratory, 950 N. Cherry Ave., Tucson, AZ 85719, USA}

\author[0000-0001-6251-0573]{Jacqueline K. Faherty}
\affiliation{Department of Astrophysics, American Museum of Natural History, Central Park West at 79th Street, New York, NY 10024, USA}

\author[0000-0003-4269-260X]{J. Davy Kirkpatrick}
\affiliation{IPAC, Mail Code 100-22, California Institute of Technology, 1200 E. California Blvd., Pasadena, CA 91125, USA}

\author[0000-0002-6294-5937]{Adam C. Schneider}
\affiliation{School of Earth and Space Exploration, Arizona State University, Tempe, AZ 85282, USA}

\author[0000-0001-7896-5791]{Dan Caselden}
\affiliation{Gigamon Applied Threat Research, 619 Western Ave., Suite 200, Seattle, WA 98104, USA}

\author[0000-0002-2592-9612]{Jonathan Gagn\'e}
\affiliation{Institute for Research on Exoplanets, Universit\'e de Montr\'eal, 2900 Boulevard \'Edouard-Montpetit Montr\'eal, QC Canada H3T 1J4}
\affiliation{Plan\'etarium Rio Tinto Alcan, Espace pour la Vie, 4801 av. Pierre-de Coubertin, Montr\'eal, Qu\'ebec, Canada}

\author[0000-0002-2387-5489]{Marc J. Kuchner}
\affiliation{NASA Goddard Space Flight Center, Exoplanets and Stellar Astrophysics Laboratory, Code 667, Greenbelt, MD 20771, USA}

\author[0000-0002-6523-9536]{Adam J. Burgasser}
\affiliation{Center for Astrophysics and Space Science, University of California San Diego, La Jolla, CA 92093, USA}

\author[0000-0003-2478-0120]{Sarah L. Casewell}
\affiliation{Department of Physics and Astronomy, University of Leicester, University Road, Leicester LE1 7RH, UK}

\author[0000-0002-1783-8817]{John H. Debes}
\affiliation{ESA for AURA, Space Telescope Science Institute, 3700 San Martin Drive, Baltimore, MD 21218, USA}

\author[0000-0003-3506-5667]{\'Etienne Artigau}
\affiliation{Institute for Research on Exoplanets, Universit\'e de Montr\'eal, 2900 Boulevard \'Edouard-Montpetit Montr\'eal, QC Canada H3T 1J4}

\author[0000-0001-8170-7072]{Daniella C. Bardalez Gagliuffi}
\affiliation{Department of Astrophysics, American Museum of Natural History, Central Park West at 79th Street, NY 10024, USA}

\author[0000-0002-9632-9382]{Sarah E. Logsdon}
\affiliation{NSF's National Optical-Infrared Astronomy Research Laboratory, 950 N. Cherry Ave., Tucson, AZ 85719, USA}

\author[0000-0003-2102-3159]{Rocio Kiman}
\affiliation{Department of Astrophysics, American Museum of Natural History, Central Park West at 79th Street, New York, NY 10024, USA}
\affiliation{Department of Physics, Graduate Center, City University of New York, 365 5th Ave., New York, NY 10016, USA}

\author[0000-0003-0580-7244]{Katelyn Allers}
\affiliation{Physics and Astronomy Department, Bucknell University, 701 Moore Ave., Lewisburg, PA 17837, USA}

\author[0000-0002-5370-7494]{Chih-chun Hsu}
\affiliation{Center for Astrophysics and Space Science, University of California San Diego, La Jolla, CA 92093, USA}

\author[0000-0001-9209-1808]{John P. Wisniewski}
\affiliation{Homer L. Dodge Department of Physics and Astronomy, University of Oklahoma, 440 W. Brooks Street, Norman, OK 73019, USA}

\author{Michaela B. Allen}
\affiliation{NASA Goddard Space Flight Center, Exoplanets and Stellar Astrophysics Laboratory, Code 667, Greenbelt, MD 20771, USA}

\author{Paul Beaulieu}
\affiliation{Backyard Worlds: Planet 9}

\author[0000-0002-7630-1243]{Guillaume Colin}
\affiliation{Backyard Worlds: Planet 9}

\author[0000-0002-4143-2550]{Hugo A. Durantini Luca}
\affiliation{IATE-OAC, Universidad Nacional de C\'ordoba-CONICET, Laprida 854, X5000 BGR, C\'ordoba, Argentina}

\author[0000-0003-2236-2320]{Sam Goodman}
\affiliation{Backyard Worlds: Planet 9}

\author[0000-0002-8960-4964]{L\'eopold Gramaize}
\affiliation{Backyard Worlds: Planet 9}

\author[0000-0002-7389-2092]{Leslie K. Hamlet}
\affiliation{Backyard Worlds: Planet 9}

\author[0000-0002-4733-4927]{Ken Hinckley}
\affiliation{Backyard Worlds: Planet 9}

\author[0000-0001-8662-1622]{Frank Kiwy}
\affiliation{Backyard Worlds: Planet 9}

\author{David W. Martin}
\affiliation{Backyard Worlds: Planet 9}

\author{William Pendrill}
\affiliation{Backyard Worlds: Planet 9}

\author[0000-0003-4083-9962]{Austin Rothermich}
\affiliation{Physics Department, University Of Central Florida, 4000 Central Florida Boulevard, Orlando, FL 32816, USA}

\author[0000-0003-4864-5484]{Arttu Sainio}
\affiliation{Backyard Worlds: Planet 9}

\author[0000-0002-7587-7195]{J\"{o}rg Sch\"umann}
\affiliation{Backyard Worlds: Planet 9}

\author{Nikolaj Stevnbak Andersen}
\affiliation{Backyard Worlds: Planet 9}

\author{Christopher Tanner}
\affiliation{Backyard Worlds: Planet 9}

\author{Vinod Thakur}
\affiliation{Backyard Worlds: Planet 9}

\author[0000-0001-5284-9231]{Melina Th\'evenot}
\affiliation{Backyard Worlds: Planet 9}

\author{Jim Walla}
\affiliation{Backyard Worlds: Planet 9}

\author{Zbigniew W\k{e}dracki}
\affiliation{Backyard Worlds: Planet 9}

\author{Christian Aganze}
\affiliation{Center for Astrophysics and Space Science, University of California San Diego, La Jolla, CA 92093, USA}

\author{Roman Gerasimov}
\affiliation{Center for Astrophysics and Space Science, University of California San Diego, La Jolla, CA 92093, USA}

\author[0000-0002-9807-5435]{Christopher Theissen}
\altaffiliation{NASA Sagan Fellow}
\affiliation{Center for Astrophysics and Space Science, University of California San Diego, La Jolla, CA 92093, USA}

\author{The Backyard Worlds: Planet 9 Collaboration}

\begin{abstract}

We present \textit{Spitzer} follow-up imaging of 95 candidate extremely cold brown dwarfs discovered by the Backyard Worlds: Planet 9 citizen science project, which uses visually perceived motion in multi-epoch \textit{WISE} images to identify previously unrecognized substellar neighbors to the Sun. We measure \textit{Spitzer} [3.6]$-$[4.5] color to phototype our brown dwarf candidates, with an emphasis on pinpointing the coldest and closest Y dwarfs within our sample. The combination of \textit{WISE} and \textit{Spitzer} astrometry provides quantitative confirmation of the transverse motion of 75 of our discoveries. Nine of our motion-confirmed objects have best-fit linear motions larger than 1$''$/yr; our fastest-moving discovery is WISEA J155349.96+693355.2 ($\mu \approx 2.15''$/yr), a possible T type subdwarf. We also report a newly discovered wide-separation ($\sim$400 AU) T8 comoving companion to the white dwarf LSPM J0055+5948 (the fourth such system to be found), plus a candidate late T  companion to the white dwarf LSR J0002+6357 at 5.5$'$ projected separation ($\sim$8,700 AU if associated). Among our motion-confirmed targets, five have \textit{Spitzer} colors most consistent with spectral type Y. Four of these five have exceptionally red \textit{Spitzer} colors suggesting types of Y1 or later, adding considerably to the small sample of known objects in this especially valuable low-temperature regime. Our Y dwarf candidates begin bridging the gap between the bulk of the Y dwarf population and the coldest known brown dwarf.

\end{abstract}

\keywords{brown dwarfs --- infrared: stars --- proper motions --- solar neighborhood}

\section{Introduction}
\label{sec:intro}

A complete census of the solar neighborhood provides the best way to identify and study the Galactic substellar population. The intrinsic faintness of the lowest temperature brown dwarfs means that we can only hope to directly image those which are nearby to the Sun. For any given substellar type, the examples most amenable to detailed follow-up observations will be those that are closest. In particular, as the \textit{James Webb Space Telescope} \citep[\textit{JWST};][]{jwst} nears launch, it is critical to identify the highest priority brown dwarf targets for spectroscopic characterization, especially pushing into the low mass/temperature regime of isolated exoplanet analogs.

By virtue of its unprecedented full-sky sensitivity at 3-5~$\mu$m, the \textit{Wide-field Infrared Survey Explorer} \citep[\textit{WISE};][]{wright10} has dramatically enhanced our ability to pinpoint the coldest brown dwarfs and reshaped our view of the solar neighborhood \citep[e.g.,][]{kirkpatrick11,cushing_y_dwarfs,luhman16ab,j0855}. The \textit{WISE} $W1$ (3.4~$\mu$m) and $W2$ (4.6~$\mu$m) bandpasses were specifically engineered to enable the selection of ultra-cold brown dwarfs via their red $W1-W2$ colors \citep{mainzer_first_brown_dwarf}. Including recent data from the NEOWISE mission \citep{neowise, neowiser}, \textit{WISE} has now surveyed the entire sky 14 times spanning almost a decade, making identification of nearby brown dwarfs based on their large apparent motions an increasingly important tool for their discovery \citep[e.g.,][]{luhman_planetx, scholz, allwise_motion_survey, allwise2_motion_survey, schneider_neowise, backyard_worlds, marocco2019, catwise_p14034}.

However, the vast \textit{WISE}/NEOWISE data set has yet to be exhaustively searched for cold and close brown dwarfs \citep[e.g.,][]{davy_parallaxes}. \textit{WISE} has detected more than 2 billion unique astronomical sources \citep{unwise_catalog}, and its full imaging archive contains over 30 trillion pixels. Despite modern computing resources, human vetting still plays an integral role in the discovery of moving objects amongst a sea of interlopers and artifacts \citep[e.g.,][]{allwise_motion_survey, allwise2_motion_survey, schneider_neowise}.

To overcome the bottleneck of visual inspection encountered by \textit{WISE}-based moving object searches, we initiated the Backyard Worlds: Planet 9 citizen science project \citep[\href{http://backyardworlds.org}{backyardworlds.org};][hereafter Backyard Worlds]{backyard_worlds}. Backyard Worlds crowdsources the visual vetting workload amongst thousands of volunteers who participate via the internet, viewing animated blinks of \textit{WISE} images spanning the full \textit{WISE}+NEOWISE time baseline. In this work we present a sample of 96 candidate extremely cold brown dwarfs discovered by Backyard Worlds volunteers. For this sample, which generally lacks high-significance detections at $W1$, we have obtained critical \textit{Spitzer}/IRAC \citep{spitzer_overview, irac} photometry at 3.6~$\mu$m (ch1) and 4.5~$\mu$m (ch2). Secure ch1 detections enable estimates of key parameters including spectral type, effective temperature, and distance.

In $\S$\ref{sec:wise} we briefly summarize the \textit{WISE} and NEOWISE missions. In $\S$\ref{sec:byw} we provide background information about the Backyard Worlds project. In $\S$\ref{sec:selection} we explain our selection of the \textit{Spitzer} follow-up targets that comprise our sample ($\S$\ref{sec:sample}). In $\S$\ref{sec:strategy} we discuss our \textit{Spitzer} observing strategy. In $\S$\ref{sec:spitzer_phot} we present our \textit{Spitzer} photometry. In $\S$\ref{sec:spectra} we present follow-up spectroscopic confirmations. In $\S$\ref{sec:astrom} we perform \textit{WISE} and \textit{Spitzer} astrometry in order to assess significance of motion. In $\S$\ref{sec:nir} we present follow-up and archival near-infrared photometry for our sample. In $\S$\ref{sec:discussion} we synthesize our photometry, spectroscopy and astrometry results, commenting on various individual objects of particular interest. We conclude in $\S$\ref{sec:conclusion}.

\section{\textit{WISE}/NEOWISE Overview}
\label{sec:wise}

\textit{WISE} is a 0.4 meter aperture space telescope in low Earth orbit, launched in late 2009. \textit{WISE} mapped the entire sky once in all four of its broad mid-infrared channels ($W1$ = 3.4~$\mu$m, $W2$ = 4.6~$\mu$m, $W3$ = 12~$\mu$m and $W4$ = 22~$\mu$m) during the first half of 2010 \citep{cutri12}. In the latter part of 2010 the two longest wavelength bands became unusable due to exhaustion of onboard cryogen. However, \textit{WISE} continued to operate in $W1$ and $W2$ through early 2011 thanks to the NEOWISE mission extension \citep{neowise}. \textit{WISE} was put into hibernation from 2011 February until 2013 December, at which point it recommenced surveying in $W1$ and $W2$ as part of the NEOWISE-Reactivation mission \citep[NEOWISE-R;][]{neowiser, cutri15}. NEOWISE-R has continued its $W1$/$W2$ observations ever since.

A typical sky location is observed for a $\sim$1 day time period once every 6 months. \textit{WISE} has now completed a total of 14 such full-sky mappings in $W1$ and $W2$. By coadding exposures within each six-monthly visit to each sky location, it is possible to construct a time series of deep and clean \textit{WISE} coadds optimized for detecting faint moving objects \citep{tr_neo2, tr_neo3, neo4_coadds} while leveraging a time baseline $>$10$\times$ longer than that of AllWISE \citep{cutri13}. The joint analysis of \textit{WISE} and NEOWISE data therefore opens up a huge discovery space for finding faint moving objects, such as cold brown dwarfs, in the mid-infrared.

\section{Backyard Worlds Overview}
\label{sec:byw}

In order to fully mine the combined \textit{WISE}+NEOWISE data set for moving object discoveries, we launched the Backyard Worlds: Planet 9 citizen science project on 2017 February 15 \citep{backyard_worlds}. Backyard Worlds crowdsources the process of visually confirming motion in \textit{WISE} images by distributing animated ``flipbooks'' via the Zooniverse web portal \citep{zooniverse}. Each flipbook shows a time-series blink covering a random\footnote{The locations are truly random; there is no pre-selection of the flipbook footprints to e.g., be centered on a suspected brown dwarf candidate.} $\sim 10' \times 10'$ patch of sky. In detail, each frame of each blink is a color-composite difference image meant to null out static background sources. Each difference image is built by creating one coadded sky pass worth of \textit{WISE} data \citep[a `time-resolved unWISE coadd';][]{tr_neo2} and then subtracting off a static sky template built by coadding  other epochs of \textit{WISE} data. Very fast-moving objects (motions of many arcseconds per year or larger) will appear unsubtracted, while slower-moving sources (down to a few hundred mas/yr) will manifest as partially subtracted `dipoles'. The blue (red) channel represents $W1$ ($W2$) so that redder (i.e., colder) moving objects will appear distinctively orange\footnote{The Backyard Worlds documentation at \url{https://www.zooniverse.org/projects/marckuchner/backyard-worlds-planet-9/about/research} contains example images illustrating this color scheme and the appearance of dipoles and fast movers in our flipbooks.}.

Although Backyard Worlds places substantial emphasis on discovering very cold and close Y dwarfs, participants are encouraged to report any moving object not presented in the prior literature \citep[i.e., not in the SIMBAD database;][]{simbad}. Examples include comoving substellar companions to higher-mass primaries \citep[e.g.,][]{wise2150}, white dwarfs with infrared excesses \citep[e.g.,][]{j0207}, cold brown dwarfs of low metallicity \citep[e.g.,][]{esdTs} and hypothesized planets in the outer solar system \citep[e.g.,][]{matese99, matese11, sheppard_trujillo, luhman_planetx, batygin_brown, p9w1, p9_3pi}.

As of 2020 April, Backyard Worlds counts more than 58,000 registered users, with the number of unique participants (including those not registered) estimated to be several times larger. Among these participants are $\sim$200 advanced users who collaborate closely with one another and the science team. Advanced users have created a number of custom motion search tools which expand the Backyard Worlds effort significantly beyond its presence on the Zooniverse web portal \citep[e.g., WiseView;][]{wiseview} and often make use of catalog querying interfaces such as SIMBAD, IRSA, Vizier \citep{vizier} and the Astro Data Lab \citep{data_lab_spie}. See Figure 1 of \cite{backyard_worlds} for an example Backyard Worlds flipbook image, and Figure 1 of \cite{catwise_p14034} for an example WiseView image sequence.

Our brown dwarf discovery and vetting processes both made extensive use of the Legacy Surveys sky viewer\footnote{\url{https://www.legacysurvey.org/viewer}}. The Legacy Surveys viewer allows for interactive exploration of maps and catalogs from \textit{WISE}, the Mayall z-band Legacy Survey \citep[MzLS;][]{mosaic3}, the Dark Energy Camera Legacy Survey \citep[DECaLS;][]{dey_overview}, the Dark Energy Survey \citep[DES;][]{des_dr1} and the Dark Energy Camera Plane Survey \citep[DECaPS;][]{decaps}. These ground-based surveys with the Blanco and Mayall 4-meter telescopes combine to provide red-optical ($z$ and $Y$ band) imaging that covers half the sky with unparalleled depth exceeding that of e.g., Pan-STARRS \citep{the_ps1_surveys} and SDSS \citep{sdss}. DES and DECaPS in particular include southern sky regions with relatively few other archival data sets available. A very faint or entirely absent counterpart in these deep red-optical surveys is suggestive of an extremely cold \textit{WISE}-detected brown dwarf. On the other hand, a diffuse red counterpart in e.g., DECaLS indicates an extragalactic source rather than a new member of the solar neighborhood. Additional catalogs of red-optical Mayall/Blanco data were accessed via the NOAO Source Catalog \citep{nsc_dr1} at Astro Data Lab. Astro Data Lab also made the unWISE Catalog \citep{unwise_catalog} accessible to Backyard Worlds citizen scientists, thereby enabling many cold brown dwarf discoveries.

\section{\textit{Spitzer} Target Selection}
\label{sec:selection}

Our \textit{Spitzer} follow-up consists of two separate observing campaigns, programs 14076 and 14299, which we will refer to as p14076 and p14299 (PI: Faherty in both cases). Target selection for both campaigns was essentially the same, though performed at different times: p14076 (p14299) targets were selected in 2018 March (2019 April).

Backyard Worlds maintains a running list of all user-submitted, previously unpublished candidate moving objects with motions that have been visually confirmed by our science team and archival photometry suggesting L, T or Y spectral types. This list currently contains $\sim$1,600 newly discovered brown dwarf candidates, and represents the parent sample from which we selected our \textit{Spitzer} photometry targets. At the time of p14076 (p14299) target selection, this parent sample of Backyard Worlds discoveries included $\sim$800 ($\sim$1,200) brown dwarf candidates.

The primary goal of our \textit{Spitzer} photometry campaigns was to pinpoint the strongest Y dwarf candidates among our moving object discoveries. As such, the majority of our targets were selected because they bear the hallmarks of potential Y dwarfs: detected only in $W2$ (undetected in $W1$ and other optical/infrared surveys) and exhibiting clear motion consistent with that of a brown dwarf in the solar neighborhood. Additionally, we selected some \textit{Spitzer} targets which have a faint $W1$ counterpart visible if any of the following three conditions were met: the crude phototype based on $W1-W2$ color suggested possible membership within the 20 pc sample; the candidate appeared to be a potential common proper motion (CPM) companion to another moving object;  the motion was exceptionally large ($\mu \gtrsim 1''$/yr). With these selection rules in mind, we re-examined our full list of Backyard Worlds discoveries prior to each \textit{Spitzer} campaign's proposal submission and requested observations for those objects which matched our criteria for meriting \textit{Spitzer} follow-up. One major driver behind our p14299 effort was the desire to phototype relatively recent Backyard Worlds discoveries before the impending retirement of \textit{Spitzer}.

We do not expect our brown dwarf target list to be contaminated by $W2$-only solar system objects. Because their apparent motions are so large, inner solar system objects are entirely nulled out by our coaddition of \textit{WISE} data into daily intervals. At Kuiper belt distances, faint solar system objects should also be largely removed by our stacking, and their degree-scale parallaxes easily distinguish them from brown dwarfs in the solar neighborhood. At distances of order 1,000 AU \citep[e.g.,][]{batygin_brown} a planetary body would remain present in our flipbooks, but the corresponding parallax of several arcminutes would again allow us to separate such a candidate from our brown dwarf targets. For parallaxes of  $\sim$1$'$ to $\sim$1$^{\circ}$, the \textit{WISE} data would appear to show one solar system body as a pair of linear tracklets at sky locations offset by roughly twice the parallax. Each of these two apparitions would be perceived during our target vetting visual inspections (performed on isolated arcminute sized cutouts) as `vanishing' during every other sky pass, which is inconsistent with the signature of a brown dwarf in the solar neighborhood. Therefore, our \textit{Spitzer} brown dwarf target sample will exclude objects at a few thousand AU or closer even though we have not attempted to perform solar system orbit linking on arcminute or degree angular scales. During our brown dwarf target selection process, sources alternating in sky position by a few arcminutes or more from one sky pass to the next would generally have been discarded and presumed to be `latent' detector artifacts, which were found to be a common contaminant in the \textit{WISE} Planet 9 search of \cite{p9w1}. At distances of $\sim$10,000-30,000 AU \citep[e.g.,][]{matese11}, the trajectory of a solar companion would be dominated by many arcseconds of oscillatory parallactic motion, but none of our \textit{Spitzer} targets display such behavior.

We ultimately selected 65 targets for p14076 and 33 targets for p14299. p14299 inadvertently re-targeted two objects from p14076 (WISEA 0651$-$8355 and WISEA 1627$-$2443), meaning that our full \textit{Spitzer} follow-up sample contains 96 unique sources. The sample presented in this work additionally includes one object (CWISE 0002+6352) for which we have analyzed serendipitous archival \textit{Spitzer} imaging from the GLIMPSE360 program \citep{glimpse}.

During p14299 target selection, we sought to avoid duplicating any objects already slated for \textit{Spitzer} observations by the CatWISE team \citep[\textit{Spitzer} program 14034, p14034 for short;][]{catwise_data_paper,catwise_p14034}. The Backyard Worlds and CatWISE moving object discovery lists overlap extensively, since both projects seek to uncover faint brown dwarfs using motions spanning the combined \textit{WISE}+NEOWISE time baseline.

\section{\textit{Spitzer} Target Sample Properties}
\label{sec:sample}

In our estimation, the combined p14076+p14299 sample is best viewed as effectively AllWISE-selected, resulting from a combination of numerous AllWISE queries issued by Backyard Worlds advanced users. 91\% (87/96) of our discoveries are present in the AllWISE catalog, despite the fact that the full \textit{WISE}+NEOWISE data set now contains $>$2$\times$ more securely detected sources than AllWISE \citep{unwise_catalog}. The discoveries of four of our nine targets not in AllWISE (WISEU 0019$-$0943, WISEU 0055+5847, WISEU 0505+3043, WISEU 2150$-$7520) are readily explained by dedicated CPM searches where \textit{WISE} visual inspection was seeded by the \textit{Gaia} DR2 \citep{gaia_mission, gaia_dr2} catalog of higher-mass objects in the solar neighborhood.

Because our \textit{Spitzer} target selection was finalized before the public release of CatWISE \citep{catwise_data_paper} and we believe our sample to be largely AllWISE-selected, we adopt AllWISE designations for our discoveries when available. In the absence of an AllWISE counterpart, we then use the CatWISE designation if one is available, and finally employ unWISE Catalog designations when neither AllWISE nor CatWISE contains a counterpart.

\begin{figure*}
\plotone{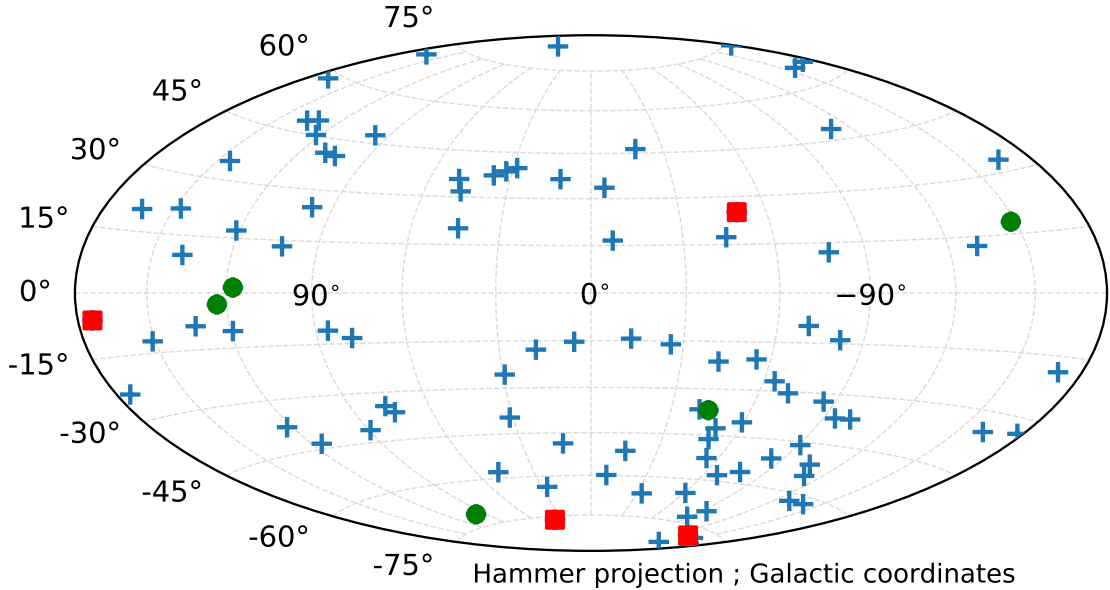}
\caption{Full-sky distribution of all 96 Backyard Worlds targets followed up with \textit{Spitzer} programs 14076 and 14299, plus one discovery (CWISE 0002+6352) with archival \textit{Spitzer} data, in Galactic Hammer projection. CPM candidates are denoted by green circles. Red squares are spurious candidates (Table \ref{tab:duds}) for which no \textit{Spitzer} counterpart was found. The locations of all other brown dwarf candidates are shown as blue plus marks.}
\label{fig:spatial_distribution}
\end{figure*}

Our sample's spatial distribution in Galactic coordinates is shown in Figure \ref{fig:spatial_distribution}. As expected, our brown dwarf candidates are scattered across the entire sky, with a significant underdensity in the crowded Galactic plane --- only 6\% (6/97) objects inhabit the $|b_{gal}| < 10^{\circ}$ sky region that accounts for 17\% of the sky. There also appears to be somewhat of an overdensity toward the south Galactic cap.

\begin{figure}
\plotone{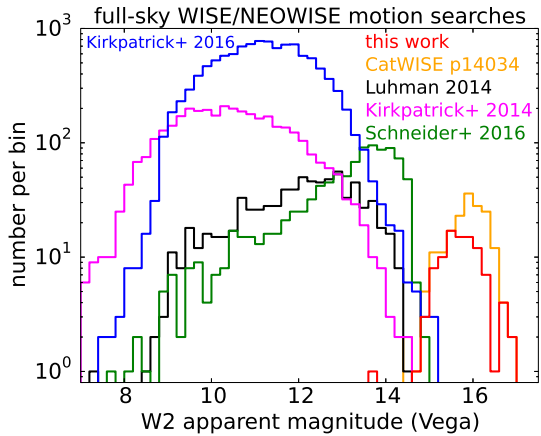}
\caption{$W2$ magnitude distributions of full-sky \textit{WISE}/NEOWISE motion survey samples (logarithmic vertical axis scale). The red histogram represents our 95 Backyard Worlds discoveries followed up with \textit{Spitzer} p14076 and p14299, plus one with archival \textit{Spitzer} data (CWISE 0002+6352). The Backyard Worlds targets are much fainter than those of all prior samples except for CatWISE \citep[\textit{Spitzer} program 14034;][]{catwise_p14034}.}
\label{fig:w2_histogram}
\end{figure}

\subsection{Targets Previously Presented by Backyard Worlds}
\label{sec:prior_papers}

The discoveries of six of our targets have been previously published by our team: WISEU 2150$-$7520 in \cite{wise2150} plus WISEA 0014+7951, WISEA 0830+2837,  WISEA 0830$-$6323, WISEA 1516+7217, and WISEA 1525+6053 in \cite{w0830}. These prior publications should be considered the definitive references for WISEU 2150$-$7520 and WISEA 0830+2837, as they contain more detailed treatments of these objects. For  WISEA 0014+7951, WISEA 0830$-$6323, WISEA 1516+7217 and WISEA 1525+6053, the present work provides significant new information by motion-confirming these targets and supplying more accurate \textit{WISE}+\textit{Spitzer} proper motion measurements for them.

\subsection{Overlap with Previously Published CatWISE \textit{Spitzer} Follow-up}
\label{sec:catwise_overlap}

Although we intended to avoid duplicating any \textit{Spitzer} follow-up of brown dwarf discoveries already published by the CatWISE team \citep{marocco2019, catwise_p14034, j1446}, we inadvertently targeted CWISEP 1359$-$4352 in p14299. We nevertheless propagate this object through our entire analysis pipeline in this work, since its p14299 \textit{Spitzer} photometry is deeper than the p14034 photometry initially presented in \cite{catwise_p14034}.

\subsection{Spurious Candidates}
\label{sec:duds}

Four brown dwarf candidates were found to have no \textit{Spitzer} counterpart (see Table \ref{tab:duds} and $\S$\ref{sec:spitzer_phot}). These spurious candidates are thought to be very faint noise excursions or artifacts in the \textit{WISE} data. In addition to the targets in Table \ref{tab:duds}, WISEA J162716.41$-$244355.4 clearly has an extended morphology based on our \textit{Spitzer} imaging and is therefore presumed to be a piece of nebulosity in the $\rho$ Ophiuchi molecular cloud complex rather than a moving object.

\subsection{\textit{WISE} Photometry}
\label{sec:wise_phot}

Table \ref{tab:wise_spitzer_photom} provides \textit{WISE} photometry for our entire sample of 96 targets. In gathering \textit{WISE} photometry, we gave preference to CatWISE photometry when available, since CatWISE fluxes are fit to linearly moving profiles that account for source motion and CatWISE incorporates $4\times$ more input $W1$/$W2$ imaging than AllWISE. In the absence of CatWISE photometry, we next checked AllWISE for photometry. In the event that neither CatWISE nor AllWISE photometry was available, we use photometry from the unWISE Catalog \citep{unwise_catalog}. Note that the unWISE Catalog performs source detection and photometry independently in $W1$ and $W2$, and so a few objects in Table \ref{tab:wise_spitzer_photom} have $W2$ photometry from the unWISE Catalog but no corresponding $W1$ photometry. Also note that in general the identifier prefixes we have chosen (typically WISEA for AllWISE) do not correspond to the origin of the \textit{WISE} photometry listed in Table \ref{tab:wise_spitzer_photom} (typically CatWISE). For WISEU 2150$-$7520 we report the custom \textit{WISE} photometry from \cite{wise2150}. Table \ref{tab:wise_spitzer_photom} lists the origin of WISE photometry reported for each target. All magnitudes quoted throughout this paper are in the Vega system unless otherwise noted.

Figure \ref{fig:w2_histogram} shows a histogram of the $W2$ magnitudes of the 96 new discoveries in our sample. The median $W2$ magnitude of this sample is $W2$ = 15.70 mag. This is much fainter than all prior full-sky \textit{WISE} motion searches except for \cite{catwise_p14034}, which had a median $W2$ magnitude of $W2$ = 15.93 mag. Both Backyard Worlds and CatWISE have been able to push much fainter than previous full-sky \textit{WISE} motion searches because they make use of time-resolved unWISE coadd images \citep{tr_neo2, tr_neo3, neo4_coadds} spanning the full \textit{WISE}+NEOWISE baseline.

\section{\textit{Spitzer} Observing Strategy}
\label{sec:strategy}

The primary goal of our \textit{Spitzer} follow-up is to obtain photometric spectral type estimates, which we typically cannot determine from \textit{WISE} because our targets generally lack secure $W1$ detections. \textit{Spitzer} ch1$-$ch2 color tends to increase monotonically toward later spectral types, and so we can use the \cite{davy_parallaxes} relation for spectral type as a function of ch1$-$ch2 color to phototype our brown dwarf candidates.

We adopted a simplistic observing strategy whereby every target is observed using the exact same \textit{Spitzer} dithering sequence --- no scaling of the total exposure time (or number of dithers) per target was performed based on anticipated \textit{Spitzer} brightness/color. We opted for a per-target Astronomical Observation Request (AOR) consisting of a ``Spiral16'' dither pattern with ``medium'' scale and a 30 second frame time per channel per dither.

The total exposure time per target in ch1 was engineered to ensure that our \textit{Spitzer} photometry can distinguish between late T and Y dwarfs, even for our faintest targets. The \textit{Spitzer} color `boundary' between late T and Y dwarfs occurs at ch1$-$ch2 $\approx$ 2.4 mag \citep{davy_parallaxes}. In our observation planning, we assume that ch2 $\approx$ $W2$, as is typical for mid-T and later brown dwarfs  \citep{davy_parallaxes}. The faintest $W2$ magnitude among our targets is $W2$ = 16.78, so we budget to achieve a ch1 signal-to-noise ratio (SNR) of 5 near the T/Y boundary according to ch1(SNR = 5) = 16.78 + 2.4 = 19.18. Our past \textit{Spitzer} observing campaigns (e.g., program 70062) have indicated that obtaining 5 dithers with 30 second frame times achieves SNR = 5 at ch1 = 18.75. Assuming that depth scales with the square root of the number of dithers, 16 dithers would then go 0.63 mags deeper at 5$\sigma$, yielding an expected 5$\sigma$ ch1 threshold of 19.38 mag, exceeding that desired with a margin of 0.2 mag. This margin can help accommodate depth variations due to factors such as zodiacal background or diffuse emission near the Galactic plane. Our targets have median ch2 $\approx$ 15.6 and median ch1$-$ch2 $\approx$ 1.9 (Table \ref{tab:wise_spitzer_photom}), so we typically obtain a ch1 uncertainty of $\sim$0.04-0.05 mag (see $\S$\ref{sec:spitzer_phot}), corresponding to ch1 SNR of $\sim$20-25.

\section{Spitzer Photometry}
\label{sec:spitzer_phot}

\textit{Spitzer} photometry is performed according to $\S$5.1 of \cite{davy_parallaxes}. In brief, we construct a custom mosaic  for each target in each of ch1 and ch2 using the MOsaicker and Point Source EXtractor \citep[MOPEX;][]{mopex} software package. For a few targets, we specially remove one or a small number of dithers where the ch1 counterpart's immediate vicinity happened to be corrupted by a cosmic ray or glint. We then perform source detection and photometry with MOPEX/APEX \citep{mopex_extraction}. We tabulate both point response function (PRF) and aperture photometry for extracted sources, and apply the appropriate aperture correction to our aperture fluxes to obtain their equivalent total fluxes. Table \ref{tab:wise_spitzer_photom} lists our final \textit{Spitzer} photometry results, where we have averaged each target's PRF and aperture photometry quantities to arrive at the quoted values. In the case of CWISE 0002+6352, we simply quote the PRF photometry, since the aperture photometry is contaminated by a nearby source.

We adopted a default source detection threshold of SNR = 5, but two exceptionally red targets required decreasing this threshold to SNR = 2: WISEU 0503$-$5648 and WISEA 1930$-$2059.

WISEU 2245$-$4333 is severely blended with a brighter neighboring source at our \textit{Spitzer} epoch, such that we are only able to extract photometry in ch2 but not in ch1. Unfortunately this means that we cannot obtain a \textit{Spitzer}-based phototype for this source, nor estimates of any other derived quantities that require a ch1$-$ch2 color measurement (e.g., photometric distance and effective temperature).

\section{Spectroscopic Follow-Up}
\label{sec:spectra}

\subsection{Magellan FIRE Spectroscopy}
We used the 6.5m Baade Magellan telescope and the Folded-port InfraRed Echellette (FIRE; \citealt{Simcoe13}) spectrograph to obtain near-infrared spectra of eight objects on our \textit{Spitzer} programs (see Table~\ref{tab:spectroscopy}). Observations were made for four sources on  2018 December 1 under clear conditions and an additional four sources under poor conditions: one object on 2019 December 10 and three objects on 2019 December 12.  For all observations, we used the prism mode and the 0.6$\arcsec$ slit  (resolution $\lambda$/$\Delta \lambda \sim$ 100) covering the full 0.8-2.5$\micron$ band. We observed all objects using a standard ABBA nod pattern with an exposure time of 120 seconds per nod. We observed an A0 standard star after each target for telluric corrections and obtained a Ne Ar lamp spectrum for wavelength calibration.  At the start of the night we used quartz lamps as domeflats in order to calibrate pixel-to-pixel response.  Data were reduced using the FIREHOSE package which is based on the MASE and Spextool reduction packages (\citealt{Bochanski09}, \citealt{Cushing04}, \citealt{vacca03}). 

\begin{figure*}
\plotone{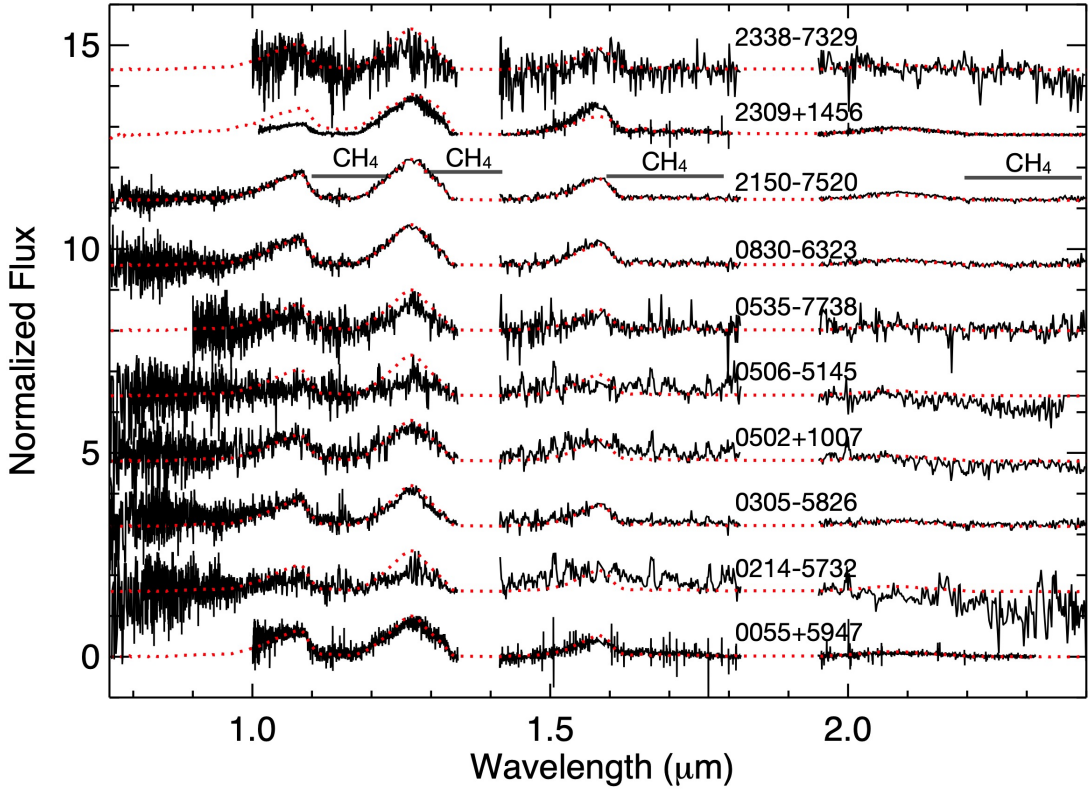}
\caption{All spectra obtained for this sample (both FIRE and NIRES) plotted in order of decreasing RA. The quality of the data varies by conditions at the telescope (see Table~\ref{tab:spectroscopy}) and faintness of each source. Overplotted on each is the T8 standard 2MASSI
J0415195$-$093506 (red curve) from \citet{Burgasser04}. All sources are +/-1 subtype from a T8 therefore it is a comparison for all.  Regions impacted by telluric absorption are removed (at $\sim$1.4~$\mu$m and $\sim$1.9~$\mu$m).}
\label{fig:spectra}
\end{figure*}

\subsection{Keck NIRES Spectroscopy}
We used the Keck II telescope and the Near-Infrared Echellette Spectrometer (NIRES; \citealt{Wilson2004}) on 2019 October 28 and 2018 September 1 to obtain 0.94-2.45~$\mu$m near infrared spectra of two objects (see Table~\ref{tab:spectroscopy}). All components on NIRES are fixed with an 0.55$''$ slit producing resolution $\sim$ 2,700 data. All sources were identified using the $K$ band finder camera and observed in an ABBA nod pattern.  Exposure times and AB frames acquired are listed in Table~\ref{tab:spectroscopy}. The data were reduced using a modified version of Spextool (\citealt{Cushing04}; see also $\S$4.4 of \citealt{kirkpatrick11}), following the standard procedure. Wavelength calibration was determined using telluric lines.  The spectra of each A-B pair were extracted individually and then combined with the other extracted pairs.  The telluric correction procedure was carried out as detailed in \citet{vacca03}.

\subsection{Spectra Results}
We show all spectra obtained for this $Spitzer$ sample in Figure~\ref{fig:spectra}.  Objects are arranged in order of decreasing right ascension from top to bottom.  Several objects were observed under poor weather conditions (see Table~\ref{tab:spectroscopy}) and therefore the quality of their data is diminished, especially at $H$ and $K$ band (e.g., WISEA J021420.18$-$573245.1).  Regardless, all objects show clear CH$_{4}$ absorption in $J$ band and we can confirm all of them to be late T dwarfs.  We compared each object against the T dwarf spectral standards in the SpeX prism library and list our best fit type in Table~\ref{tab:spectroscopy}.  

For uniform comparison in Figure~\ref{fig:spectra}, we overplot the T8 standard 2MASSI J0415195$-$093506 (red curve) from \citet{Burgasser04} on each object. All sources are classified as T8, within 1 subtype of T8, or broadly defined as a late T dwarf and therefore our standard choice serves as a robust comparative object.  Several sources are high S/N ($>$ 10 across the wavelength coverage) and well fit to the T8 standard (within 1 spectral type).  These include WISEU J215018.46$-$752053.0 which was the subject of \citet{wise2150} as it was discovered to be comoving with a \textit{Gaia} DR2 L1 dwarf,  WISEA J083019.97$-$632305.4 which has \textit{Hubble Space Telescope} (\textit{HST}) photometry reported in \cite{w0830} consistent with a late T dwarf, and WISEU J005559.88+594745.0 which is a new comoving companion reported in this paper (see $\S$\ref{sec:0055}). WISEA J083019.97$-$632305.4 and WISEU J005559.88+594745.0 both have $J$, $H$,and $K$ band spectra with clear and strong CH$_{4}$ absorption consistent with the standard implying they are field sources with no unusual parameters.

In the case of WISEA J021420.18$-$573245.1, WISEA J050238.28+100750.0, and WISEA J233816.47$-$732929.7 the spectra are too noisy to ascertain an exact type.  Each of these sources is notably later than T6.  In the case of WISEA J050615.56$-$514521.3, WISEA J053512.01$-$773829.7, and WISEA J083019.97$-$632305.4 these sources are all at least T8 but possibly later. Finally, WISEA J230930.58+145633.1 is best fit by the T8 standard.  This source along with WISE J00559.88+594745.0 were collected using the NIRES instrument on Keck II.  Unlike the FIRE prism spectra shown for all other sources, the Keck NIRES spectra are stitched together from cross dispersed orders.  This can (although not always) lead to poorly fit continuum shapes where order overlaps were deprecated by low signal to noise sources. WISEA J230930.58+145633.1 appears to have slightly depleted $J$ band flux and enhanced $H$ and $K$ band flux however this may be due to stitching in the merged spectrum of the source.  A band-by-band comparison to the standard is more compatible and therefore we do not conclude that this object is necessarily peculiar.

\section{Astrometry}
\label{sec:astrom}

\subsection{Astrometric Analysis Overview}
\label{sec:astrom_intro}

Because our brown dwarf candidates were selected based on visually perceived source movement, we need a means to quantify our confidence in the motion of each object. This can be accomplished by checking whether or not each target's joint \textit{WISE}+\textit{Spitzer} astrometry falls along a consistent linear trajectory. The combination of \textit{Spitzer} and \textit{WISE} data points is critical, since our \textit{Spitzer} follow-up imaging is completely independent of any anomalous occurrence in the \textit{WISE} data which may have misled our target selection efforts (e.g., noise, artifacts, blending, imperfect blink registration).

Our targets are (by selection) better detected at 4-5~$\mu$m than at 3-4~$\mu$m, so we only perform the detailed astrometric analyses of $\S$\ref{sec:wise_astrom}-\ref{sec:pm_fits} in $W2$ and ch2. The basic steps of our astrometric analyses are: obtaining \textit{Gaia}-calibrated $W2$ and ch2 positions spanning the $\sim$2010-2019 time period ($\S$\ref{sec:wise_astrom}, $\S$\ref{sec:spitzer_astrom}) and then fitting these with a linear motion model ($\S$\ref{sec:pm_fits}). The linear trajectory fitting results can then be used to assess significance of motion as described in $\S$\ref{sec:pm_fits}. As a byproduct, we also obtain more accurate motion measurements than would be possible using \textit{WISE} data alone.

\subsection{\textit{WISE} Astrometry}
\label{sec:wise_astrom}

Full details of our $W2$ astrometry procedure are provided in $\S$8.3 of \cite{catwise_p14034}. Our \textit{WISE} astrometry is based on catalogs constructed by running the \verb|crowdsource| pipeline \citep{decaps, unwise_catalog} on unWISE coadds. These unWISE coadds time-slice the available $W2$ data in a number of ways, for example: six-monthly intervals, yearly intervals, the entire pre-hibernation time period, and the entire post-hibernation time period. Detailed definitions of our unWISE time slices for $W2$ astrometry are provided in Table 5 of \cite{catwise_p14034}. For each target, we create $W2$ coadds and catalogs corresponding to all of these different time slices, then choose a subset of the available $W2$ detections that best covers the entire \textit{WISE}/NEOWISE time period while never double-counting any $W2$ observations. The most common choice is simply a pair of $W2$ astrometric detections: one for the entire pre-hibernation time period coadded together and another for the entire post-hibernation time period coadded together. In the present work we incorporate the sixth year of \textit{WISE} data (2017 December - 2018 December), whereas \cite{catwise_p14034} only used five years of \textit{WISE} data.

The unWISE coadds simply propagate the native exposure-level (``Level 1b'') world coordinate system (WCS), which can have systematics at the 100-200 mas level\footnote{\url{http://wise2.ipac.caltech.edu/docs/release/allwise/expsup/sec5_2b.html}}. We therefore recalibrate our $W2$ astrometry to \textit{Gaia} DR2 accounting for the \textit{Gaia} calibrator proper motions. Note that none of our brown dwarf candidates are detected by \textit{Gaia} --- rather, our astrometry is tied to that of brighter/warmer nearby stars present in \textit{Gaia}. We typically achieve a $W2$ versus \textit{Gaia} bright end scatter of 40-45 mas per coordinate, just $\sim$1/60 of the \textit{WISE} pixel sidelength. The complete set of $W2$ (RA, Dec) coordinates used in our \textit{WISE}+\textit{Spitzer} linear motion fits is provided in Table \ref{tab:wise_positions}.

\subsection{Spitzer Astrometry}
\label{sec:spitzer_astrom}

Our \textit{Spitzer} astrometry methodology is as described in $\S$8.4 of \cite{catwise_p14034}; here we provide only a concise summary. Our ch2 astrometry is measured from mosaics rather than single-dither \textit{Spitzer} frames and proceeds from the same set of extracted catalogs used for our ch2 photometry ($\S$\ref{sec:spitzer_phot}). We do not simply adopt the native WCS of the mosaics, which is inherited from that of the single-dither \textit{Spitzer} frames tied to 2MASS, since these have astrometry accurate only at the order hundred mas level \citep{martin_spitzer_astrometry}. We can achieve much better astrometric accuracy of $\sim$20-25 mas per coordinate by recalibrating the mosaic WCS to \textit{Gaia} \citep{martin_spitzer_astrometry, catwise_p14034}. In detail, we match the ch2 mosaic extractions to \textit{Gaia} DR2 using the native WCS, then refit six astrometric parameters per mosaic: the WCS CRPIX coordinates (i.e., translations along the two sky directions) and all four CD matrix elements (i.e., scale factors governing the mosaic pixel size and orientation)\footnote{See \cite{wcs} for precise definitions of these WCS parameters.}. Our fitting minimizes residuals between the WCS-predicted pixel locations of the \textit{Gaia} DR2 sources (propagated to the relevant \textit{Spitzer} epoch according to their proper motions) and the measured pixel coordinate centroids. The median per-coordinate astrometric scatter relative to \textit{Gaia} after our mosaic WCS recalibration is 20 mas.

The complete set of ch2 (RA, Dec) coordinates used in our \textit{WISE}+\textit{Spitzer} linear motion fits is provided in Table \ref{tab:spitzer_positions}. The ``method number'' column of Table \ref{tab:spitzer_positions} is an integer code indicating which set of \textit{Gaia} DR2 calibrator selection cuts was applied during astrometric recalibration, with these codes defined in Table 6 of \cite{catwise_p14034}. The $N_{calib}$ column lists the number of \textit{Gaia}-\textit{Spitzer} calibrators employed for each target's mosaic. We sought to always obtain at least 5 astrometric calibrators per mosaic, and only failed to achieve this in one case: WISEA 1627$-$2443. The lack of calibrators in this field is due to heavy dust extinction within the \textit{Gaia} bandpass, and in any event WISEA 1627$-$2443 itself turns out to be an extended piece of nebulosity.

\subsection{Linear Motion Fits}
\label{sec:pm_fits}

We combine our $W2$ and ch2 astrometry by performing linear fits of the motion along both the RA and Dec directions. We do not attempt to include a parallactic motion component in these fits. The fits are performed using weighted linear least squares, where the per-coordinate uncertainties are those listed in Table \ref{tab:wise_positions} and \ref{tab:spitzer_positions}. The least squares fitting propagates these positional uncertainties into formal uncertainties on the RA and Dec direction linear motions. No outlier rejection is performed, since we generally have only a handful of astrometric data points per target, and all $W2$ and ch2 detections used for motion fitting were individually vetted in advance. The best-fit linear motions and their uncertainties are provided in Table \ref{tab:wise_spitzer_pm} for all targets detected in our \textit{Spitzer} ch2 imaging.

A primary aim of our \textit{WISE}+\textit{Spitzer} astrometric analysis is to determine quantitatively whether each target is indeed moving. To do this, we adopt the same significance of motion metric and threshold as in \cite{catwise_p14034}: $\chi^2_{motion} > 23.01$, where $\chi^2_{motion} = \mu^2_{\alpha}/\sigma^2_{\mu_{\alpha}} + \mu^2_{\delta}/\sigma^2_{\mu_{\delta}}$. This threshold corresponds to a $< 10^{-5}$ probability for random scatter in the linear motion measurement to have induced such a large $\chi^2_{motion}$ value for an object that is actually stationary.

\begin{figure}
\plotone{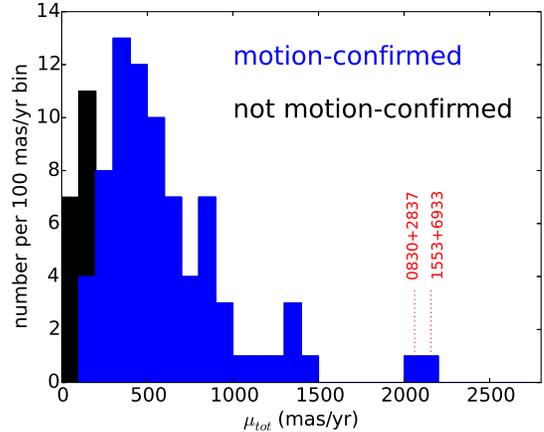}
\caption{Distribution of best-fit total linear motions for our brown dwarf candidates. The number of sample members per bin that were not motion-confirmed is shown in black, stacked on top of the number of motion-confirmed targets per bin (blue). Objects with $\mu_{tot} \gtrsim 170$ mas/yr, corresponding to approximately half of a \textit{WISE} pixel of displacement over the available time baseline, tend to be motion-confirmed. One new discovery (WISEA 1553+6933) has best-fit $\mu_{tot} > 2000$ mas/yr, and eight have $\mu_{tot} > 1000$ mas/yr; these number counts do not include WISEA 0830+2837 ($\mu_{tot} \approx 2060$ mas/yr), since this object's discovery was previously presented in \cite{w0830}.}
\label{fig:total_motion_hist}
\end{figure}

Figure \ref{fig:total_motion_hist} shows a histogram of the best-fit total linear motions obtained. Our $\chi^2_{motion}$ threshold tends to be exceeded for $\mu_{tot} \gtrsim$ 170 mas/yr at the typical $W2$ magnitude of our sample. 79 of our 97 sample members are motion-confirmed according to our $\chi^2_{motion}$ criterion. One, CWISEP 1359$-$4352, was previously motion-confirmed in \cite{catwise_p14034}. In another two cases, WISEA 1627$-$2443 and WISEA 1350$-$8302, we vetoed the motion confirmation after visual inspection of the \textit{Spitzer} imaging. WISEA 1627$-$2443 is seen to be spatially extended at \textit{Spitzer} resolution, and thus our measured positions likely exhibit excess scatter relative to our positional uncertainties that assume pointlike profiles. This source is presumed to be a piece of nebulosity rather than a moving object. Upon more detailed examination of the $W2$ and ch2 images of WISEA 1350$-$8302, we determined that the \textit{WISE} brown dwarf candidate is actually a very faint $\sim$8$''$ separation pair of \textit{Spitzer} sources, the redder of which we have listed in Table \ref{tab:spitzer_positions}. Therefore, the fitted motion is a spurious artifact of comparing the \textit{WISE} centroid, which effectively averages the positions of two faint \textit{Spitzer} sources, against the ch2 position of one of those \textit{Spitzer} objects.

Since we use the same motion confirmation approach and significance threshold as did the CatWISE team for their p14034 \textit{Spitzer} follow-up \citep{catwise_p14034}, it is interesting to compare our fraction of successful motion confirmations to theirs. 64\% (114/177) of CatWISE p14034 brown dwarf candidates were motion confirmed versus 79\% (77/97) for our Backyard Worlds \textit{Spitzer} sample. The higher rate of confirmations for Backyard Worlds likely arises because the CatWISE p14034 sample is fainter by $\sim$0.2-0.3 magnitudes in the median, leading to correspondingly noisier \textit{WISE} astrometry.

\section{Near Infrared Photometry}
\label{sec:nir}

\subsection{Mont-M\'egantic/CPAPIR Follow-up}
\label{sec:omm}

We observed 42 of our brown dwarf candidates in the MKO $J$ band with the Cam\'era PAnoramique Proche Infra-Rouge \citep[CPAPIR;][]{cpapir} wide-field (30$'$ $\times$ 30$'$) near-infrared camera located at the Observatoire du Mont-M\'egantic (OMM) 1.6\,m observatory \citep{omm} between 2018 May and 2020 January. 10 of these OMM targets were well-detected ($> 4\sigma$), 11 were marginally detected ($< 4\sigma$) and 21 were not detected.

Targets were always observed near the center of the lower-right quadrant of the CPAPIR camera because the amplifier of the upper-right quadrant is malfunctioning. The data were reduced with a custom Python pipeline that was standardized for CPAPIR/OMM data. Images are flat-fielded with a standard white field inside the telescope dome observed at the start or end of the night. A sky image is built from a median of the observations, taken across small, random dithering patterns and where all sources at the 2MASS positions are masked. The sky image is subtracted from each individual frame, and a linear astrometric solution is obtained with a local implementation of the \url{astrometry.net} tool \citep{astrometry_dot_net} anchored on index files built from \textit{Gaia} DR2 data \citep{gaia_dr2, lindegren_gaia_dr2}. Individual images are then median combined into a final frame, and a new astrometric solution is built from \url{astrometry.net} and \textit{Gaia} DR2 including second-degree distortions as Simple Imaging Polynomial (SIP) polynomials.

Aperture photometry was calculated for each target with a custom IDL pipeline using an aperture twice as large as the FWHM reported by the OMM data reduction pipeline (1.7$''$--4.9$''$ with a median of 2.1$''$), centered at the position predicted from the \textit{WISE}+\textit{Spitzer} motion solutions. Using twice the FWHM ensures that small deviations in predicted versus observed positions do not affect the photometric measurement, without needing to estimate a new centroid for very faint targets. The background sky was estimated from the median of an annulus with inner and outer radii 3 and 6 times as large as the photometric aperture, respectively.

Photometric zero points were estimated by measuring the flux of all 2MASS catalog entries within the field of view (while avoiding the outermost 250 pixel edges) and calculating the median zero point that translates from CPAPIR $J$ band fluxes to 2MASS $J$ band magnitudes. Limiting 5$\sigma$ depths were calculated for non-detections by sampling a thousand uniform-random positions in the field of view while avoiding 2MASS entries. The average of the absolute flux values at 84\% and 16\% percentiles of the random distribution multiplied by 5 was then transformed to a 5$\sigma$ detection limit in magnitudes with the appropriate zero point. The 5$\sigma$ depths ranged from 15.7 to 19.2\,mag with a median value of 18.1\,mag.

\subsection{Archival Near Infrared Photometry}

We have also compiled archival near infrared (NIR) photometry for members of our sample using the WFCAM Science Archive \citep[UKIRT/WFCAM data;][]{wsa} and VISTA Science Archive \citep[VISTA/VIRCAM data;][]{vsa}. We issued a cone search in the vicinity of each target and visually vetted  possible NIR counterparts to avoid spurious positional cross-matches. We searched for archival NIR photometry in the $Y$, $J$, $H$, and $K$ bands ($K_{S}$ rather than $K$ in the case of VIRCAM). The NIR counterparts are predominantly drawn from the UKIRT Hemisphere Survey in the north \citep[UHS;][]{uhs} and the VISTA Hemisphere Survey in the south \citep[VHS;][]{vhs}, though VIKING \citep{viking} and UKIDSS \citep{ukidss} also contribute. By selection, all members of our sample lack 2MASS \citep{tmass} counterparts.

\subsection{Merged Near Infrared Photometry}

Table \ref{tab:jhk_phot} provides a merged compilation of OMM and archival near infrared photometry for motion-confirmed targets, in the $YJHK/K_{S}$ bands. In cases where both OMM and archival follow-up $J$ band photometry are available, Table \ref{tab:jhk_phot} quotes the higher significance measurement. Table \ref{tab:omm_extra} reports additional OMM follow-up results for targets that are not motion-confirmed and/or have higher significance $J$ band measurements in archival data sets. Our follow-up OMM photometry agrees well with archival photometry in cases where both are available.

\begin{figure*}
\plotone{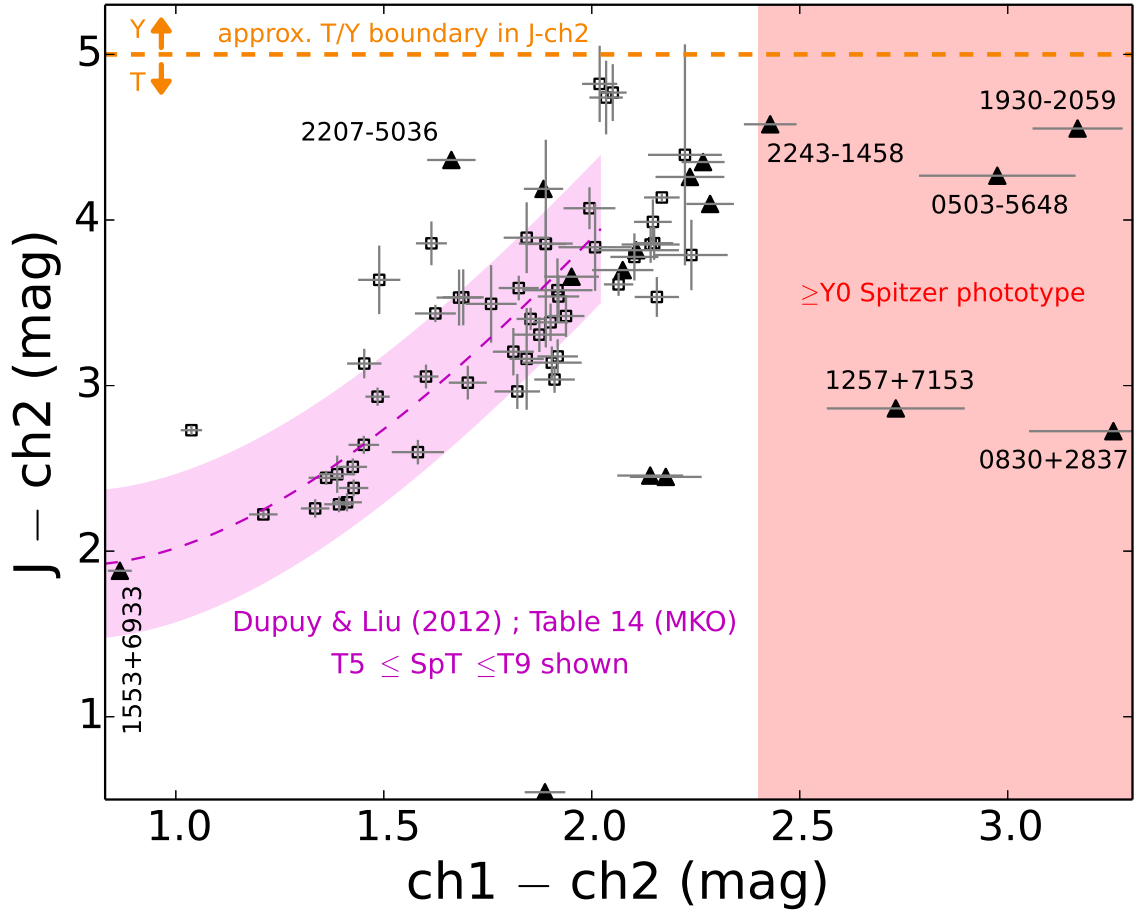}
\caption{$J - $ch2 versus ch1$-$ch2 for motion-confirmed targets with both colors available. Follow-up $J$ band photometry comes from OMM/CPAPIR and archival $J$ band photometry is obtained from the WFCAM/VISTA archives. Black squares represent detections. All $J -$ch2 lower limits (black triangles) are based on 5$\sigma$ $J$ limits. The dashed purple line shows the relation for mid-late T dwarfs from \cite{dupuy_liu_2012}, with the purple shaded region indicating the 1$\sigma$ scatter in that relation. Our motion-confirmed sample generally follows the expected trend. The light red shaded region denotes the \textit{Spitzer} color range within which our phototyping procedure yields $\ge$Y0 estimates, ch1$-$ch2 $\geq$ 2.4 mag. The dashed orange horizontal line indicates the approximate boundary between late T and early Y dwarfs in terms of $J-$ch2 color. Our $J-$ch2 limit for WISEAR 2207$-$5036, a suspected subdwarf due to its high kinematics, indicates that it is also a color outlier.}
\label{fig:j_minus_ch2}
\end{figure*}

Figure \ref{fig:j_minus_ch2} shows a color-color plot of $J-$ch2 versus ch1$-$ch2 for motion-confirmed targets with both colors available. The data points are by and large in good agreement with the expected trend for brown dwarfs \citep{dupuy_liu_2012}, providing further assurance that our motion-confirmation process has weeded out static contaminants which might display rather discrepant colors. WISEAR 2207$-$5036, a suspected subdwarf due to its high kinematics ($\S$\ref{sec:subdwarfs}), stands out as likely the strongest color outlier. Even if all other sample members are typical field brown dwarfs, we would statistically expect some fraction to scatter $> 1\sigma$ off of the \cite{dupuy_liu_2012} locus; this likely explains most or all of the Figure \ref{fig:j_minus_ch2} $J$ band detections falling modestly outside of the purple shaded region. Analogous color-color diagrams for $Y-$ch2, $H-$ch2, and $K-$ch2 again show agreement with the expected trends for brown dwarfs, though we have omitted those plots due their paucity of NIR detections.

In addition to the ground-based NIR photometry gathered here, five of our targets have \textit{HST} follow-up NIR photometry \citep{w0830}. Owing to the non-standard bandpasses of these \textit{HST} data, we have chosen not to merge the \textit{HST} measurements with our ground-based photometry in this work. 

\section{Results \& Discussion}
\label{sec:discussion}

\subsection{Derived Parameters}
\label{sec:derived_params}

\begin{figure}
\plotone{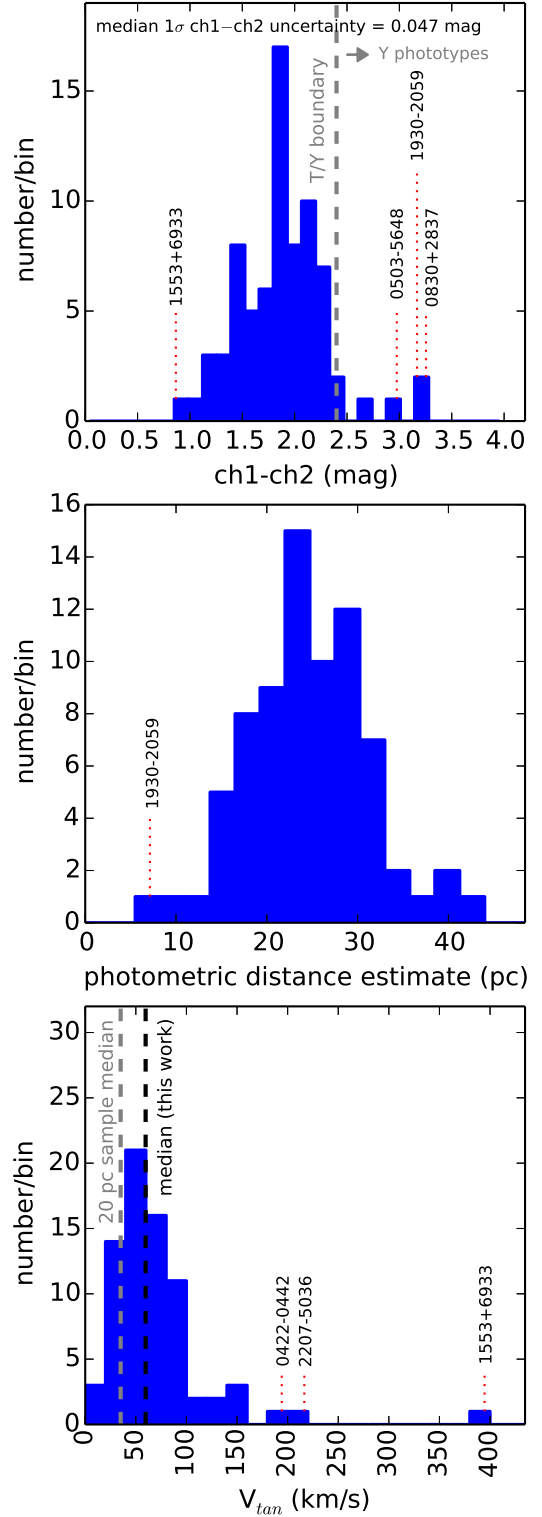}
\caption{Distributions of \textit{Spitzer} colors (top), photometric distance estimates (middle) and $V_{tan}$  from Table \ref{tab:derived_properties} for motion-confirmed targets whose spectral types could be estimated based on our \textit{Spitzer} photometry. Short names of selected extreme objects are included as annotations.}
\label{fig:gold_sample_hist}
\end{figure}

Table \ref{tab:derived_properties} lists various properties that we are able to derive based on our ch1 and ch2 magnitudes, for the sample of motion-confirmed objects with both of these \textit{Spitzer} photometric data points available\footnote{WISEA 0535$-$6445 is excluded from Table \ref{tab:derived_properties} despite having a \textit{Spitzer} color measurement, since its ch1$-$ch2 $\approx$ 0 color does not place a strong constraint on its spectral type or absolute magnitude. See $\S$\ref{sec:wise_0535} for further discussion of this object.}. These derived parameters are:

\begin{itemize}

\item Phototype: We estimate phototypes using the \cite{davy_parallaxes} relation for spectral type as a function of ch1$-$ch2 color. We use the same phototyping procedure as \cite{catwise_p14034}, which contains a more detailed explanation of our methodology and associated caveats. In brief, \cite{catwise_p14034} estimate an RMS phototyping uncertainty of approximately $\pm$1 subtype. Larger errors between phototype and real spectral type for individual objects are of course possible, and photometric type estimates should never be considered substitutes for true spectral types. For the four objects successfully assigned a true spectral type and subtype in Table \ref{tab:spectroscopy}, our phototypes achieve the expected level of agreement: an RMS difference of 1.03 subtype.

\item Absolute ch2 magnitude: We estimate the absolute ch2 magnitude using the \cite{davy_parallaxes} relation for $M_{ch2}$ as a function of ch1$-$ch2. Again, our procedure is the same as that of \cite{catwise_p14034}, which contains further details, particularly in relation to the quoted uncertainties.

\item Distance: Photometric distance estimates for members of our sample follow from the combination of each object's measured apparent ch2 magnitude and $M_{ch2}$ estimate. Details of how we quote these distances and their associated uncertainties can be found in \cite{catwise_p14034}.

\item Effective temperature: We provide $T_{\rm eff}$ estimates based on the \cite{davy_parallaxes} relation for effective temperature as a function of ch1$-$ch2 color. No $T_{\rm eff}$ value is listed if ch1$-$ch2 $<$ 0.9, since the \cite{davy_parallaxes} $T_{\rm eff}$(ch1$-$ch2) relation is not applicable in that \textit{Spitzer} color regime.

\item Tangential velocity: $V_{tan}$ estimates follow from the combination of our $\mu_{tot}$ values in Table \ref{tab:wise_spitzer_pm} and our photometric distance estimates. As shown in Figure \ref{fig:gold_sample_hist}, our set of targets appears to have a higher typical $V_{tan}$ value than that found for the \cite{davy_parallaxes} volume-limited sample of mid-late T and Y dwarfs within 20 pc. The 180 \cite{davy_parallaxes} objects within 20 pc have a median $V_{tan}$ of 35 km/s, whereas our sample's median $V_{tan}$ is 60 km/s. This presumably owes to a bias whereby our motion-selected sample preferentially contains relatively faint and fast-moving targets.

\end{itemize}

Table \ref{tab:derived_properties} also includes ch2 reduced proper motions\footnote{Defined as $H_{ch2} = m_{ch2} + 5$log$_{10}\mu + 5$.}, calculated by combining linear motion fitting results from Table \ref{tab:wise_spitzer_pm} with the apparent ch2 magnitudes from Table \ref{tab:wise_spitzer_photom}. Figure \ref{fig:gold_sample_hist} shows histograms of ch1$-$ch2 color, photometric distance and $V_{tan}$ for objects in Table \ref{tab:derived_properties}. Further discussion subsections draw heavily on the Table \ref{tab:derived_properties} parameters to highlight notable aspects and members of our sample.

\subsection{The 10 pc and 20 pc Samples}
\label{sec:nearest}

Studies of the substellar mass function are currently limited by incompleteness of the local ($\lesssim 20$ pc) brown dwarf sample at the latest spectral types \citep{davy_parallaxes}. A core objective of this work was therefore to help pinpoint previously unidentified members of the 20 pc census. 

In terms of our nearest motion-confirmed discoveries, two have best-fit photometric distance estimates (Table \ref{tab:derived_properties}) within 10 pc: WISEA 0830+2837 and WISEA 1930$-$2059. WISEA 1930$-$2059 also has a 1$\sigma$ distance upper envelope value within 10 pc. For WISEA 0830+2837, \cite{w0830} have published a \textit{WISE}+\textit{Spitzer} trigonometric parallax of 90.6 $\pm$ 13.7 mas placing this object most likely outside of the 10 pc sample, though still possibly closer than 10 pc within the 1$\sigma$ parallax uncertainty. WISEU 0503$-$5648 has a central distance estimate larger than 10 pc but could still be closer than 10 pc within its 1$\sigma$ photometric distance estimate uncertainty.

Sixteen of our motion-confirmed discoveries have central photometric distance estimates within 20 pc. An additional fifteen of our motion-confirmed discoveries have central photometric distance estimates larger than 20 pc but still within $1\sigma$ of the 20 pc threshold. Considering that the \cite{davy_parallaxes} 20 pc sample of late T and Y dwarfs consisted of 235 objects (46 of which have central parallax values placing them within 10 pc), our discoveries represent a sizable new batch of candidate 20 pc sample members.

\subsection{Y Dwarf Candidates}
\label{sec:y_dwarfs}

Given our use of the \cite{davy_parallaxes} phototyping relation, we have effectively placed our threshold for Y dwarf candidacy at ch1$-$ch2 $>$ 2.4 magnitudes. Five of our motion-confirmed brown dwarf candidates have best-fit \textit{Spitzer} colors most consistent with $\ge$Y0 phototypes: WISEU 0503$-$5648, WISEA 0830+2837, WISEA 1257+7153, WISEA 1930$-$2059 and WISEA 2243$-$1458. WISEA 0830+2837 has already been discussed thoroughly in \cite{w0830}, and is thought to be of type $\ge$Y1. Of the other four motion-confirmed Y dwarf candidates first presented in this work, three have exceptionally red \textit{Spitzer} colors most consistent with types $\ge$Y1. This new crop of $\ge$Y1 candidates discovered by Backyard Worlds is particularly significant given that only 4-6 spectroscopically confirmed brown dwarfs are presently known in this regime: WISE 0350$-$5658 \citep[Y1;][]{kirkpatrick12}, WISE 0647$-$6232 \citep[Y1;][]{kirkpatrick12}, WISE 1541$-$2250 \citep[Y1;][]{cushing_y_dwarfs}, WISE 2354$-$0240 \citep[Y1;][]{schneider_hst_spectra}, WISE 1828+2650  \citep[$\ge$Y2 with highly uncertain $T_{\rm eff}$;][]{beichman_w1828} and WISE 0855$-$0714 \citep[L and M band spectra;][]{w0855_m_band, w0855_l_band}. The impact of our sample's likely coldest four discoveries in terms of mass function analyses is also considerable: only nine previously known objects in the \cite{davy_parallaxes} 20 pc compilation inhabit the corresponding $T_{\rm eff} < 400$ K effective temperature range.

\begin{figure*}
\plotone{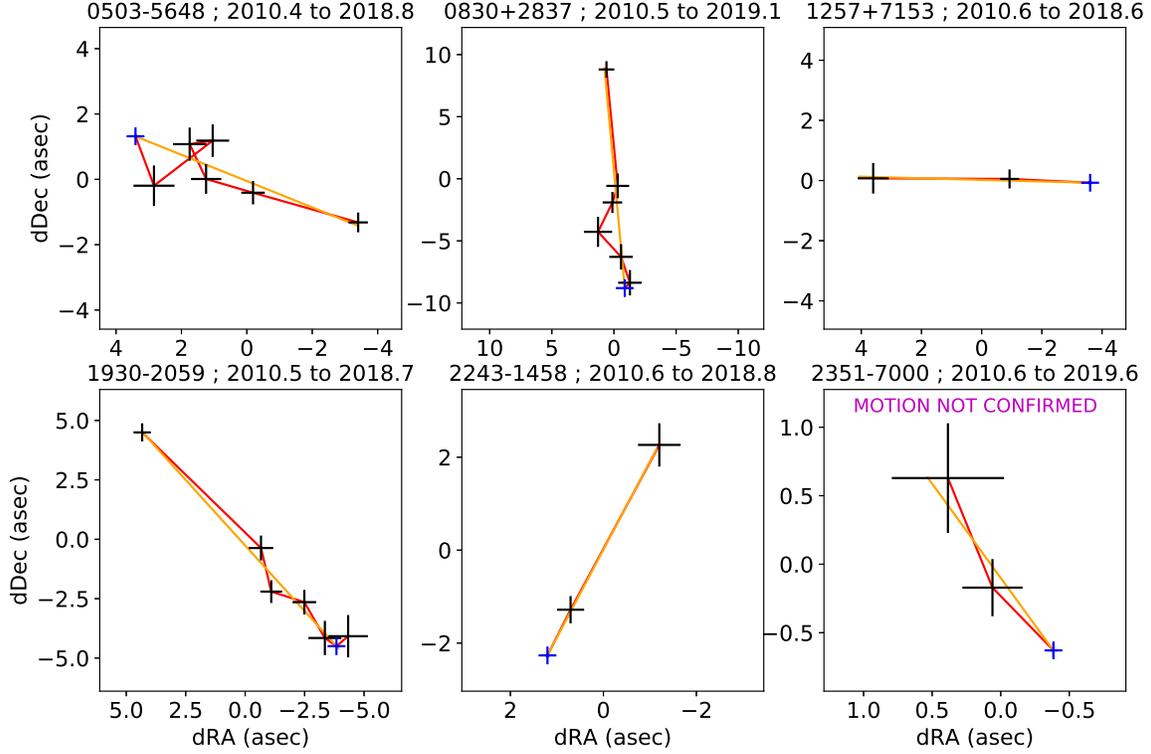}
\caption{Combined \textit{WISE}+\textit{Spitzer} astrometry for our six targets with best-fit ch1$-$ch2 $>$ 2.4 mag, corresponding to phototypes $\ge$Y0. Each panel's title lists the short object identifier and time period spanned by its astrometry. All but WISEA 2351$-$7000 are motion-confirmed. The WISEA 2351$-$7000 trajectory appears plausibly linear, but its significance of motion falls just short of exceeding our $\chi^2_{motion}$ threshold for motion confirmation. dRA (dDec) is the positional offset along the RA (Dec) direction relative to the midway point between the minimum and maximum RA (Dec) for each object. dRA is in units of angular separation, rather than being a simple difference of RA coordinate values. Black plus marks represent the \textit{WISE} $W2$ locations and their $\pm$1$\sigma$ uncertainties. \textit{Spitzer} astrometry is represented by a blue  plus mark of arbitrary size in each panel; the actual \textit{Spitzer} uncertainties are much smaller than these symbols and would be difficult to perceive if plotted to scale. Red line segments follow the astrometric measurements in a time-ordered fashion. The best-fit linear motion solutions (Table \ref{tab:wise_spitzer_pm}) are shown as orange lines and cover the same time period spanned by the combined set of \textit{WISE} and \textit{Spitzer} detections. As illustrated by the relatively large size of the black plus marks, the time series of measured \textit{WISE} positions can be quite noisy. For example, the loop-like phenomenon seen for WISEU 0503$-$5648 results from centroid measurement noise, not parallactic motion.}
\label{fig:radec_plots}
\end{figure*}

Of our five motion-confirmed discoveries with phototypes $\ge$Y0, four are more than 2$\sigma$ redder than our T/Y threshold color of ch1$-$ch2 = 2.4 mag. On the other hand, WISEA 2243$-$1458 is within its $1\sigma$ color uncertainty of being classified as a late T dwarf. A sixth member of our sample (WISEA 2351$-$7000) would have a Y phototype based on its ch1$-$ch2 color, but falls just short of exceeding our $\chi^2_{motion}$ significance threshold (Table \ref{tab:wise_spitzer_pm}). Figure \ref{fig:radec_plots} shows \textit{WISE}+\textit{Spitzer} motion trajectories for all six targets with ch1$-$ch2 $>$ 2.4 mag, including WISEA 2351$-$7000. The WISEA 2351$-$7000 \textit{WISE}+\textit{Spitzer} astrometric trajectory looks plausibly linear, but more data will be needed to conclusively establish whether this source is moving.

Each of our five motion-confirmed Y dwarf candidates is labeled in Figure \ref{fig:j_minus_ch2}. Clearly, the existing $J$ limits for these source are not deep enough to confirm that their $J-$ch2 colors are securely in the $J-$ch2 $\gtrsim$ 5 mag regime occupied by Y dwarfs. Therefore, deeper ground-based NIR imaging of our Y dwarf candidates would be of high value, with the exception of WISEA 0830+2837, which already has extreme $F125W-$ch2 $>$ 9.36 mag and $F105W-$ch2 $>$ 9.56 mag limits available \citep{w0830}.

\subsection{Absence of Discoveries as Cold as \\ WISE 0855$-$0714}
\label{sec:w0855}

Although WISEA 0830+2837 may potentially be similar to WISE 0855$-$0714 (the coldest known brown dwarf) in terms of ch1$-$ch2 and $F125W-$ch2 colors, WISEA 0830+2837 is $\sim$1.5 magnitudes more luminous at ch2 \citep{w0830} and thus presumably warmer. Therefore, recent \textit{Spitzer} follow-up from Backyard Worlds and CatWISE appears not to have identified even one object as cold as or colder than WISE 0855$-$0714 \citep[$T_{\rm eff} \approx 250 $ K;][]{j0855}. This may seem surprising given that motion surveys like Backyard Worlds and CatWISE are pushing much deeper than the relatively bright WISE 0855$-$0714 magnitude of $W2 \approx 14$ (see Figure \ref{fig:w2_histogram}). \cite{wright_0855} argued that a complete search of the \textit{WISE} data to $W2 = 16$ should find between 3 (16$^{th}$ percentile) and 34 (84$^{th}$ percentile) additional objects\footnote{These numbers are lower than those quoted in \cite{wright_0855} by 1 because the \cite{wright_0855} values include WISE 0855$-$0714 itself.} comparable to WISE 0855$-$0714, with a median of 14. The Backyard Worlds and CatWISE \textit{Spitzer} target lists both have complex selection functions because they each amalgamate results from numerous contributing searches, but it is unlikely that either (or their union) is currently complete to $W2 = 16$ (see especially $\S$3.3 of \citealt{catwise_data_paper} regarding CatWISE Preliminary incompleteness). Adjusting the \cite{wright_0855} prediction to the typical magnitude of our Backyard Worlds sample ($W2 \approx 15.7$; see $\S$\ref{sec:sample}) yields a corresponding 16$^{th}$ to 84$^{th}$ percentile range of 1.6 to 22 WISE 0855$-$0714 analogs, with a median of 9.

We find it conceivable that a small number of WISE 0855$-$0714 analogs have been detected by \textit{WISE}/NEOWISE but have so far not been pinpointed by either Backyard Worlds or CatWISE. For instance, the combined Backyard Worlds and CatWISE motion discovery lists total $\sim$4,000 objects, yet fewer than 300 of these were imaged with \textit{Spitzer}. One cannot predict which \textit{WISE} sources will turn out to be reddest in \textit{Spitzer} based on faint $W2$ detections alone, so it is certainly possible that the parent Backyard Worlds plus CatWISE motion discovery samples contain new WISE 0855$-$0714 analogs not yet recognized as such due to lack of \textit{Spitzer} follow-up. This potential incompleteness affects both Backyard Worlds and CatWISE, and could easily arise from e.g., blending of an extremely cold brown dwarf with one or more background contaminants, yielding a spuriously blue $W1-W2$ color. Another selection bias specific to CatWISE is that very faint and fast-moving sources may not appear in the catalog at all, since CatWISE source detection is performed on static sky coadds spanning more than half a decade \citep{catwise_data_paper}, significantly diluting the appearance of objects moving faster than $\sim$1$''$/yr.

To summarize, the absence of additional objects as cold as WISE 0855$-$0714 suggests that their true abundance is probably not on the higher side of imprecise estimates based purely on the nearby location of WISE 0855$-$0714 (i.e., there likely are not of order ten such objects with $W2 < 15.7$). On the other hand, the absence of such discoveries in the Backyard Worlds and CatWISE \textit{Spitzer} follow-up samples (zero identified versus 1.6 expected at the distribution’s 16$^{th}$ percentile) does not definitively establish an underabundance relative to the \cite{wright_0855} forecast given the extremely low number statistics and possible incompleteness. An updated 20 pc census of brown dwarfs that incorporates recent Backyard Worlds and CatWISE discoveries can provide improved space density and mass function estimates, thereby enabling a more rigorous assessment of whether the close-by detection of WISE 0855$-$0714 is anomalous given its as yet unrivaled temperature. That analysis is beyond the scope of this work, and will be presented in a dedicated follow-on paper (Kirkpatrick et al., in prep.).

\subsection{Objects with Largest Motions}
\label{sec:fastest}

\subsubsection{Subdwarf Candidates}
\label{sec:subdwarfs}

Few mid-T or later subdwarfs are presently known, so identifying more such examples is a critical step toward developing an understanding of very cold substellar objects at low metallicity \citep[e.g.,][]{zhang_sdt}. High tangential velocity is a potential indicator of a relatively old object with low metallicity possibly belonging to the Milky Way thick disk or halo. A threshold value for large $V_{tan}$ sometimes employed in solar neighborhood studies is 100 km/s \citep[e.g.,][]{faherty_bdkp}. Ten objects in Table \ref{tab:derived_properties} have central $V_{tan}$ estimates $> 100$ km/s. To highlight only the most extreme of our targets in terms of tangential velocity as potential subdwarfs, we examine the set of motion-confirmed targets with estimated $V_{tan} > 150$ km/s: WISEA 0422$-$0442 ($V_{tan} = 194^{+31}_{-27}$ km/s), WISEA 1553+6933 ($V_{tan} = 395^{+48}_{-44}$ km/s) and WISEAR 2207$-$5036 ($V_{tan} = 216^{+36}_{-32}$ km/s). Mid-late T subdwarfs may have unusually red $J - $ch2 colors relative to their ch1$-$ch2 colors \citep[e.g.,][]{wolf1130}, motivating a check of whether available NIR photometry indicates that our high $V_{tan}$ outliers are also color outliers. As shown in Figure \ref{fig:j_minus_ch2}, WISEAR 2207$-$5036 has a substantially redder $J-$ch2 color limit than would be expected for a normal brown dwarf with its ch1$-$ch2 $= 1.66 \pm 0.06$ color, bolstering its subdwarf candidacy. On the other hand, WISEA 0422$-$0442 has an NIR detection ($J = 19.43 \pm 0.23$), yielding a $J - $ch2 = $3.49 \pm 0.23$ color that is consistent with its T8 phototype.

WISEA 1553+6933 stands apart from the rest of our brown dwarf candidates due to its exceedingly high tangential velocity estimate. WISEA 1553+6933 is also our sample's fastest-moving source in terms of best-fit total linear motion ($\mu > 2''$/yr). Despite not being especially faint by the standards of our sample ($\textrm{ch2} \approx 15.5$), it was missed by prior searches due to severe blending with a static background source during pre-hibernation \textit{WISE} observations. Our OMM follow-up establishes a limit of $J > $ 17.34 for WISEA 1553+6933, corresponding to a $J - $ch2 $>$ 1.88 color constraint. This color limit is not sufficiently stringent to rule out the possibility that WISEA 1553+6933 is a normal mid-T brown dwarf. WISEA 1553+6933 merits additional follow-up as a candidate mid-late T type subdwarf, including deeper $J$ band imaging.

\subsubsection{Reduced Proper Motion}
\label{sec:reduced_pm}

Reduced proper motion can be a valuable tool for identifying low-luminosity sources, such as Y dwarfs and subdwarfs, in the absence of trigonometric parallaxes. Since our sample lacks trigonometric parallaxes, we cannot directly compute absolute ch2 magnitudes to pinpoint our most intrinsically faint sources. Reduced proper motion replaces parallax in the absolute magnitude formula with total proper motion, on the premise that large apparent motion tends to indicate that a source is relatively nearby. Thus, we can single out objects of especially low luminosity within our sample (independent of absolute magnitude versus color relations) based on their large reduced proper motions.

\begin{figure}
\plotone{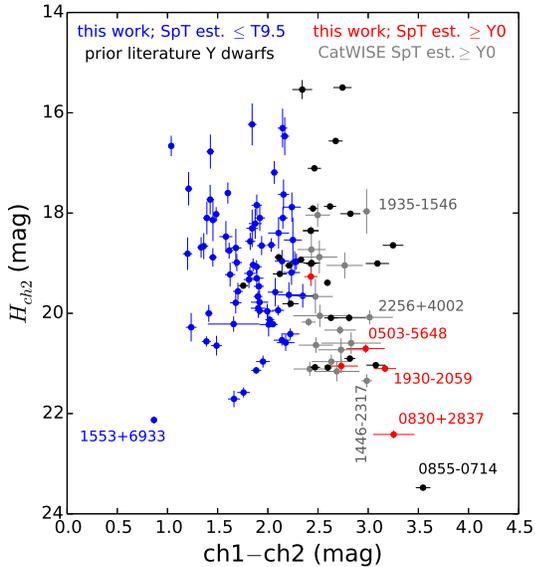}
\caption{Reduced proper motion diagram showing all Y dwarfs from the prior literature (black) and all motion-confirmed targets from this study with \textit{Spitzer} phototypes available. Our targets with best-fit ch1$-$ch2 color most consistent with spectral type Y are shown in red, while all of our objects with earlier spectral type estimates are shown in blue. Recent CatWISE discoveries with \textit{Spitzer}-based Y phototypes are also plotted in gray \citep{catwise_p14034}. Our three discoveries with largest central ch1$-$ch2 colors are individually labeled (red annotations), as are the three reddest CatWISE discoveries \citep[gray annotations;][]{marocco2019, catwise_p14034, j1446}.}
\label{fig:reduced_pm}
\end{figure}

Figure \ref{fig:reduced_pm} shows the reduced proper motions of our sample as a function of ch1$-$ch2 color. In this reduced proper motion diagram, our sample's reddest objects (by ch1$-$ch2 color) are beginning to bridge a previously wide gap between the coldest known brown dwarf \citep[WISE 0855$-$0714;][]{j0855} and the rest of the Y dwarf population. WISEA 0830+2837 in particular stands out as inhabiting this formerly empty region of parameter space, and has the largest reduced proper motion of any member of our sample ($H_{ch2}$ = $22.42 \pm 0.07$ mag). Note that WISEA 0830+2837 is indeed alone in the gap between WISE 0855$-$0714 and other Y dwarfs. CWISEP 1446$-$2317 previously fell in a similar region of parameter space, but its ch1$-$ch2 color has recently been revised substantially blueward relative to that initially presented in \cite{catwise_p14034}. Still, the WISEA 0830+2837 reduced proper motion is lower than that of WISE 0855$-$0714 by more than a magnitude. As discussed in \cite{w0830}, WISEA 0830+2837 may represent a heretofore missing link between the bulk of the as-yet identified Y dwarfs and WISE 0855$-$0714. WISEA 1553+6933 has the second highest reduced proper motion among our sample, with $H_{ch2} = 22.13 \pm 0.06$, while also being one of our bluest targets in ch1$-$ch2 color; we suspect that WISEA 1553+6933 is a T type subdwarf on account of its high kinematics ($\S$\ref{sec:subdwarfs}).

\subsubsection{Total Linear Motion}
\label{sec:high_pm}

Nine of our motion-confirmed discoveries have best-fit total linear motions larger than $1''$/yr: WISEU 0048+2508, WISEA 0422$-$0442, WISEA 0806$-$0820, WISEA 0830+2837, WISEA 1553+6933, WISEA 1628+1604, WISEA 1930$-$2059, WISEAR 2207$-$5036 and WISEA 2245$-$4333. Of these nine, two have best-fit total linear motions larger than 2$''$/yr: WISEA 0830+2837 and WISEA 1533+6933. The suspected subdwarf WISEA 1553+6933 has the largest total linear motion among our sample, with $\mu = 2160 \pm 55$ mas/yr. Several of our discoveries with $\mu_{tot} > 1''$/yr were missed by prior brown dwarf color selections due to blending at early \textit{WISE} epochs: CWISE 0002+6352, WISEU 0048+2508, WISEA 1553+6933, WISEA 1930$-$2059, and WISEU 2245$-$4333. By visually surveying for motion, Backyard Worlds citizen scientists were able to spot these previously overlooked members of the solar neighborhood.

\subsection{Candidate Common Proper Motion Systems}
\label{sec:cpm}

Five of our targets were considered potential CPM companions to earlier type primaries upon being selected for \textit{Spitzer} follow-up, and a sixth CPM candidate (CWISE 0002+6352) has archival \textit{Spitzer} data available. The comoving pair consisting of WISEU 2150$-$7520 and its L dwarf primary has already been discussed extensively in \cite{wise2150}. WISEU 0505+3043 was targeted as a possible comoving companion to the extreme subdwarf LSPM J0505+3043 but turned out to be entirely spurious (see Table \ref{tab:duds}). The other remaining four CPM candidates are discussed in detail below.

\subsubsection{CWISE J000229.93+635217.0}

The motion of CWISE 0002+6352 appears strikingly similar to that of the $\sim$5.5$'$ distant DC white dwarf LSR J0002+6357 \citep{limoges15}, which has accurate parallax and proper motion measurements available from \textit{Gaia}. Our \textit{WISE}+\textit{Spitzer} linear motion fitting gives $\mu_{\alpha} = 802 \pm 49$ mas/yr, $\mu_{\delta} = 44 \pm 47$ mas/yr for CWISE 0002+6352, versus $\mu_{\alpha} = 918.9 \pm 0.1$ mas/yr, $\mu_{\delta} = 108.3 \pm 0.1$ mas/yr for LSR J0002+6357 \citep{gaia_dr2}. The CWISE 0002+6352 RA (Dec) motion component differs from that of LSR J0002+6357 by 2.4$\sigma$ (1.3$\sigma$). Additionally, the \textit{Gaia} parallax of LSR J0002+6357 (38.1 mas, corresponding to 26.3 pc) places it within the 1$\sigma$ photometric distance interval of CWISE 0002+6352 ($28.6^{+5.6}_{-5.1}$ pc, Table \ref{tab:derived_properties}). Restricting to this distance range, only 12 sources in the entire \textit{Gaia} DR2 catalog have both proper motion components consistent with those of CWISE 0002+6352 to within 3$\sigma$. Therefore, the probability of having one such \textit{Gaia} source land at least as close on the sky as LSR J0002+6357 does to CWISE 0002+6352 by random chance is $7.7 \times 10^{-6}$.

\begin{figure*}
\plotone{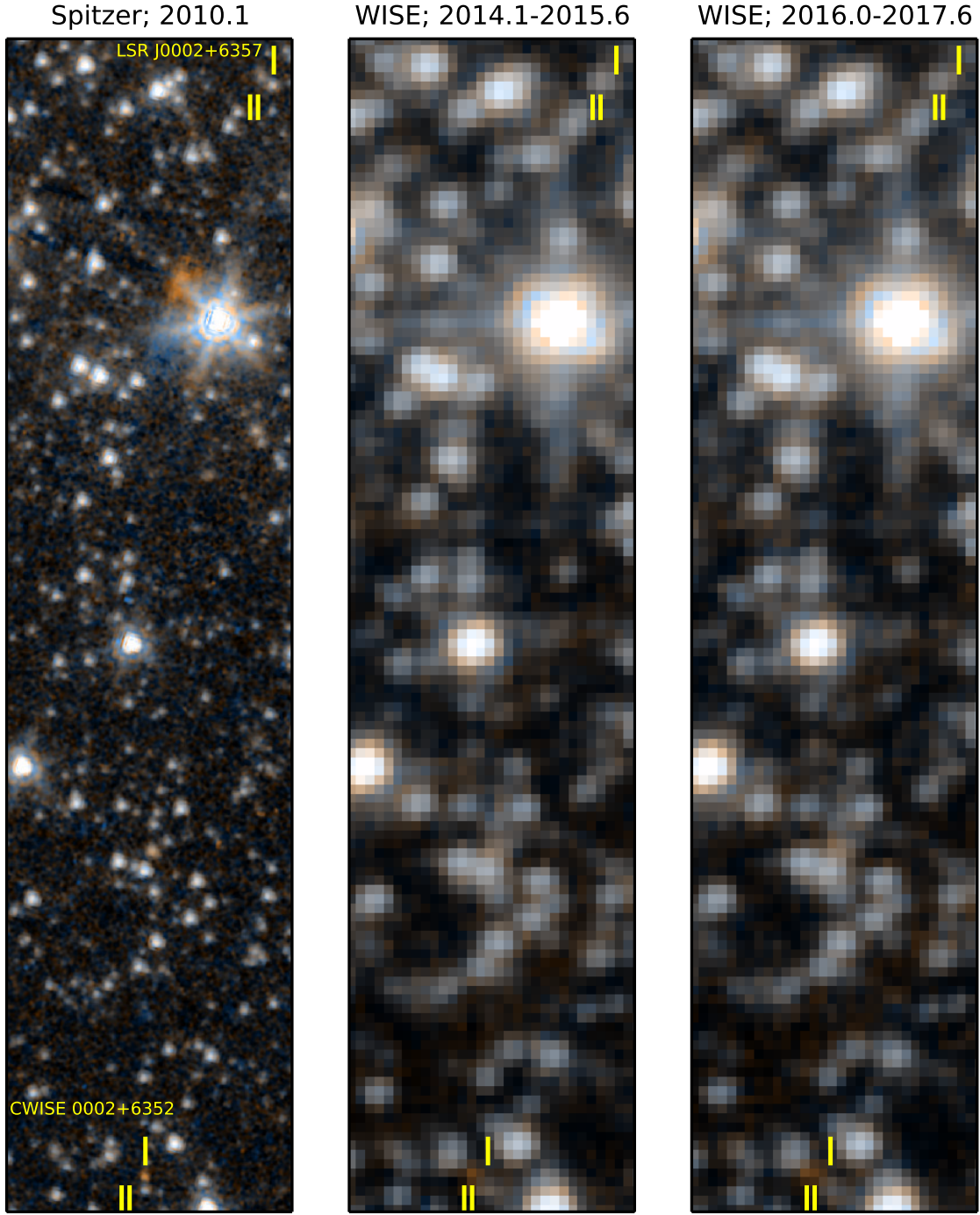}
\caption{Time series of \textit{WISE} and \textit{Spitzer} images illustrating the similar proper motions of our late T discovery CWISE 0002+6352 (bottom of each cutout) and the DC white dwarf LSR J0002+6357 (upper right of each cutout). Each panel is a two-band color composite rendering. In each case $W1$ (ch1) is represented by the blue color channel and $W2$ (ch2) is represented by the red color channel. The \textit{Spitzer} color composite at left is built from archival GLIMPSE360 imaging. \textit{WISE} images (center and right panels) are coadds spanning different portions of the post-reactivation time period. CWISE 0002+6352 appears distinctively orange in comparison to the relatively white/blue background source population that has $W1-W2$ $\approx$ ch1$-$ch2 $\approx$ 0. East is left and north is up. Yellow vertical lines track the west to east motion of this pair over the $\sim$2010-2018 time period. Each postage stamp is $1.6' \times 6.2'$ in angular extent. Follow-up NIR astrometry would help to conclusively determine whether CWISE 0002+6352 and LSR J0002+6357 are indeed physically associated by refining the former's measured proper motion.}
\label{fig:j0002}
\end{figure*}

While this false alarm probability suggests a good likelihood that CWISE 0002+6352 is a bona-fide companion to LSR J0002+6357, the moderate motion discrepancies remain concerning, particularly the difference of $> 100$ mas/yr along the RA direction. Significant blending of the brown dwarf's \textit{WISE} counterpart at essentially all epochs in this highly crowded field (see Figure \ref{fig:j0002}) means that our \textit{WISE}+\textit{Spitzer} motion may carry systematic uncertainties larger than those we quote based on statistics alone. In addition to a spectroscopic confirmation of CWISE 0002+6352 to corroborate (or discredit) its photometric distance estimate, a future NIR astrometric data point free of blending would help provide a more conclusive determination as to whether CWISE 0002+6352 is indeed physically associated with LSR J0002+6357. Based on its \textit{Spitzer} photometry, we predict $J \approx$ 18.8 for CWISE 0002+6352. CWISE 0002+6352 is not detected by Pan-STARRS \citep{the_ps1_surveys}.

If CWISE 0002+6352 and LSR J0002+6357 are physically associated, their angular separation of $330.9'' \pm 0.4''$ would translate to a projected physical separation of 8691 $\pm$ 20 AU. This would be second largest projected physical separation of any known white dwarf plus mid-late T dwarf system, following only LSPM 1459+0857AB \citep[16500-26500~AU;][]{dayjones}. LSR J0002+6357 has serendipitous \textit{Spitzer} observations from the GLIMPSE360 program, with ch1 = 15.05 $\pm$ 0.05 and ch2 = 14.91 $\pm$ 0.08, though it appears slightly blended with a neighboring source in that archival imaging (see Figure \ref{fig:j0002}).

\subsubsection{WISEU J001908.31$-$094323.3}
Based on visual inspection we recognized WISEU 0019$-$0943 to share a very similar motion with the nearby M dwarf LP 704$-$85. Our WISEU 0019$-$0943 linear motion agrees with the \textit{Gaia} DR2 proper motion of LP 704$-$85 to within 1$\sigma$ in terms of both $\mu_{\alpha}$ and $\mu_{\delta}$. WISEU 0019$-$0943 is $19.1'' \pm 0.4''$ distant from LP 704$-$85, comparing the \textit{Gaia} epoch 2015.5 position of LP 704$-$85 to that predicted at the same epoch by our linear motion model for WISEU 0019$-$0943. Using the full-sky \textit{Gaia} DR2 catalog, we find that there are only 2,601 \textit{Gaia} sources that have both $\mu_{\alpha}$ and $\mu_{\delta}$ within $1\sigma$ of the values listed for WISEU 0019$-$0943 in Table \ref{tab:wise_spitzer_pm}. This yields a chance alignment probability of $5.6 \times 10^{-6}$.

However, our \textit{Spitzer}-based photometric distance estimate for WISEU 0019$-$0943 is in some tension with the \textit{Gaia} DR2 parallax for LP 704$-$85. The WISEU 0019$-$0943 photometric distance estimate ($d$ = $31.1^{+4.6}_{-4.0}$ pc) is 3$\sigma$ discrepant from the LP 704$-$85 distance according to \textit{Gaia} DR2 ($d$ = 44.9 $\pm$ 0.2 pc). We note that this discrepancy could be eliminated if WISEU 0019$-$0943 were itself a pair of T7 brown dwarfs rather than a single T7 brown dwarf, which would then place its central distance estimate at 44.0 pc. Archival VHS images do show a pair of sources near the WISEU 0019$-$0943 location, separated by 3.4$''$. However, the WISEU 0019$-$0943 \textit{Spitzer} ch2 counterpart looks pointlike, whereas a $\sim$3$''$ separation pair of sources with similar ch2 apparent brightnesses should have yielded a significantly extended profile given that the ch2 PRF FWHM is $\sim$2$''$. We therefore believe that one of the nearby VHS NIR detections (the fainter source in $J$ band) is an unrelated background object. We consider our current data to be inconclusive regarding whether WISEU 0019$-$0943 is indeed physically associated with LP 704$-$85. If physically associated, the projected separation would be $857 \pm 16$ AU.  LP 704$-$85 has no published spectrum, but its $J_{2MASS}$ apparent magnitude and \textit{Gaia} parallax combine to give an absolute $J_{2MASS}$ magnitude of 8.9, consistent with spectral type $\sim$M4 \citep{hawley_sdss}.

\subsubsection{WISEU J005559.88+594745.0}
\label{sec:0055}

\begin{figure*}
\plotone{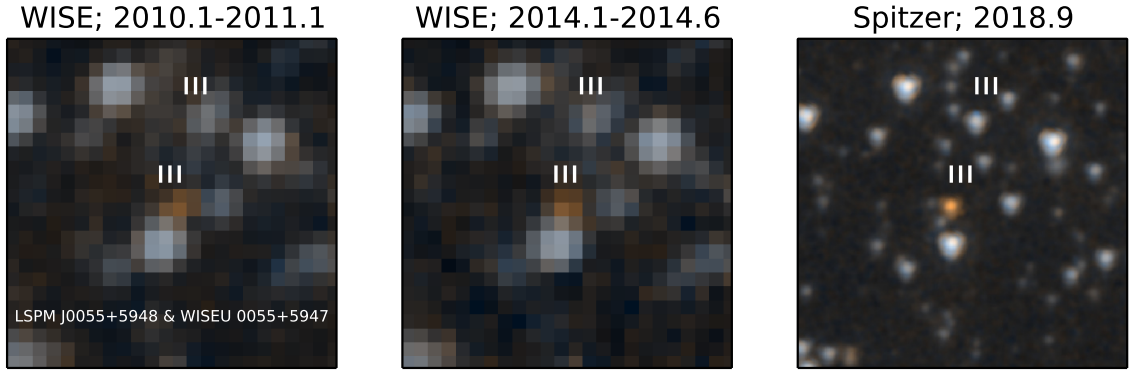}
\caption{Time series of \textit{WISE} and \textit{Spitzer} images illustrating the common proper motion of our T8 discovery WISEU 0055+5947 (center of each cutout) and the DC white dwarf LSPM J0055+5948 (upper right). Each panel is a two-band color composite rendering. In each case $W1$ (ch1) is represented by the blue color channel and $W2$ (ch2) is represented by the red color channel. \textit{WISE} images are one-year coadds spanning the pre-hibernation period (left) and first year post-reactivation (center). The \textit{Spitzer} color composite at right is built from p14076 IRAC imaging (PI: Faherty). The T8 comoving companion at center is strikingly orange in comparison to the relatively white/blue background source population that has $W1-W2$ $\approx$ ch1$-$ch2 $\approx$ 0. East is left and north is up. The white vertical lines track the west to east motion of this pair over the $\sim$2010-2019 time period. Each postage stamp is $1.1' \times 1.1'$ in angular extent.}
\label{fig:j0055}
\end{figure*}

Based on visual inspection we recognized WISEU 0055+5947 to share a very similar motion with the nearby DC white dwarf LSPM J0055+5948 \citep[see Figure \ref{fig:j0055};][]{limoges13}. Our WISEU 0055+5947 linear motion agrees with the \textit{Gaia} DR2 proper motion of LSPM J0055+5948 to within 2$\sigma$ in terms of both $\mu_{\alpha}$ and $\mu_{\delta}$. WISEU 0055+5947 is $17.6'' \pm 0.1''$ distant from LSPM J0055+5948, comparing the \textit{Gaia} epoch 2015.5 position of LSPM J0055+5948 to that predicted at the same epoch by our linear motion model for WISEU 0055+5947. Using the full-sky \textit{Gaia} DR2 catalog, we find that there are only 351 \textit{Gaia} sources that have both $\mu_{\alpha}$ and $\mu_{\delta}$ within $2\sigma$ of the values listed for WISEU 0055+5947 in Table \ref{tab:wise_spitzer_pm}. This yields a chance alignment probability of just $6.4 \times 10^{-7}$.

The WISEU 0055+5947 photometric distance estimate provides further evidence for physical association with LSPM J0055+5948. The white dwarf has a \textit{Gaia} parallax of 43.78 $\pm$ 0.07 mas, corresponding to a distance of 22.8 pc. This agrees with our \textit{Spitzer}-based photometric distance estimate for WISEU 0055+5947 ($d$ = $22.2^{+3.3}_{-2.9}$ pc; Table \ref{tab:derived_properties}) to within 1$\sigma$. Adding further corroboration, our Keck/NIRES spectral type of T8 for the secondary ($\S$\ref{sec:spectra}) agrees with our photometric estimate of T7 within the expected phototyping uncertainty ($\S$\ref{sec:derived_params}). The chance alignment probability drops further to $4.9 \times 10^{-8}$ when accounting for the consistency of the brown dwarf distance estimate with the white dwarf parallax in combination with the similarity of their proper motions. The projected physical separation between the brown dwarf and white dwarf is 402 $\pm$ 3 AU using the white dwarf's \textit{Gaia} parallax to convert from arcseconds of angular separation to projected AU. We conclude that WISEU 0055+5947 is indeed a comoving companion of LSPM J0055+5948. This makes WISEU 0055+5947 an interesting target for further benchmarking studies given the potential for determining the system's age using that of the white dwarf primary.

LSPM J0055+5948 was identified as a DC white dwarf by \citet{limoges15} and is also in the \cite{fusillo} \textit{Gaia} DR2 white dwarf catalog. According to \cite{fusillo} LSPM J0055+5948 has $T_{\rm eff} = 4623 \pm 21$ K, $log(g) = 7.82 \pm 0.02$ and $M = 0.474 \pm 0.010 M_{\odot}$ when modeled assuming a pure H atmosphere \citep{fusillo}, as suggested by \citet{limoges13}. To confirm these parameters we also obtained the Pan-STARRS $griz$ \citep{the_ps1_surveys}, 2MASS $JHK_s$, and AllWISE $W1$,$W2$ photometry of LSPM~J0055+5948 and determined T$_{\rm eff}$ and $log(g)$ using the photometric SED and \textit{Gaia} DR2 parallax. For a DA composition and fitting to the \citet{bergeron} cooling models we obtain T$_{\rm eff}$=4734 $\pm$81~K and $log(g)$ = 7.87 $\pm$ 0.07, consistent within the uncertainties with the \citet{fusillo} values.

Using the \citet{fusillo} parameters and the cooling models from \citet{bergeron} for DA white dwarfs, we determine a cooling time of 5.01$^{+0.24}_{-0.23}$~Gyr. We then use the semi-empirical initial mass-final mass relationship for white dwarfs from \citet{cummings} to determine the initial mass of the white dwarf progenitor to be 1.38$^{+0.47}_{-0.21}$~M$_{\odot}$, a likely solar-type star. Combining the progenitor mass with the MIST isochrones \citep{mesa, dotter}, we determine the total age of the system to be 10$\pm$3~Gyr. The age we determine is consistent if we use the SED determined $T_{\rm eff}$ and $log(g)$.

It should be noted that while we have assumed this DC white dwarf is likely to have a hydrogen rich atmosphere, the \citet{fusillo} modelling gives  $T_{\rm eff} = 4654 \pm 17$ K, $log(g) = 7.84 \pm 0.02$ and $M = 0.473 \pm 0.010 M_{\odot}$ when modeled assuming a pure He atmosphere, which is also consistent with the values we have determined for the white dwarf, meaning even if LSPM~J0055+5948 is a helium atmosphere white dwarf, our estimate of the system age will not change. LSPM~J0055+5948 has ch1 = $14.91 \pm 0.01$ and ch2 = $14.90 \pm 0.02$ based on our \textit{Spitzer} follow-up imaging.

Combining our total system age constraint and $T_{\rm eff}$ estimate from Table \ref{tab:derived_properties} yields a mass range of $56 \pm 9 \ M_{\rm Jup}$ for WISEU 0055+5947 based on the \cite{saumon_marley} model grids.

The WISEU 0055+5947 plus LSPM J0055+5948 pair is currently only the fourth such wide system to be discovered after LSPM 1459+0857AB \citep{dayjones}, WD0806-661AB \citep{luhman_0806} and COCONUTS-1AB \citep{coconuts}. 
The projected separation of  $\sim$400~AU for WISEU 0055+5947 is closer than the other three comoving systems, all of which have projected separations of at least 1000~AU (COCONUTS-1AB: 1280~AU, WD0806-661AB: 2500~AU, LSPM1459+0857AB: 16500-26500~AU). All four systems have T dwarf companions or later: both COCONUTS-1B and LSPM 1459+0857B are T4/5 dwarfs and WD0806-661B is a probable Y dwarf, and is almost certainly of planetary mass \citep{luhman14}. All systems are old, with ages of at least 2~Gyr, and WISEU 0055+5947 is probably the oldest, based on its low effective temperature and age estimate. The wide separation of the brown and white dwarfs means that the brown dwarf is unlikely to have been affected by the evolution of the white dwarf progenitor, and also that the white dwarf evolution is unlikely to have been truncated during a common envelope phase. This is supported by the fact that none of the white dwarfs in these systems are of low enough mass to have formed via binary evolution (e.g., \citealt{marsh}).  The fact that the components of these binaries are comoving, but have not affected each other's evolution makes them ideal benchmark systems for determining spectroscopic parameters of old brown dwarfs.

\subsubsection{WISEA J075438.20+090044.9}

Based on \textit{WISE}/NEOWISE imaging alone, the WISEA 0754+0900 motion appears visually similar to that of the 780$''$ distant M6 dwarf LP 483-66. The \textit{WISE}-based proper motion of WISEA 0754+0900 ($\mu_{\alpha} = 267 \pm 72$ mas/yr, $\mu_{\delta} = -179 \pm 90$ mas/yr according to the CatWISE 2020 catalog) is consistent with the \textit{Gaia} DR2 proper motion of LP 483-66 ($\mu_{\alpha} = 414$ mas/yr, $\mu_{\delta} = -233$ mas/yr) to within $\sim$2$\sigma$ along both directions. Our photometric distance estimate for WISEA 0754+0900 ($26.6^{+4.0}_{-3.5}$ pc; Table \ref{tab:derived_properties}) is also consistent with the LP 483-66 parallax from \textit{Gaia} (33.2 mas, corresponding to a distance of 30.1 pc). However, our \textit{WISE}+\textit{Spitzer} motion ($\mu_{\alpha} = 166 \pm 52$ mas/yr, $\mu_{\delta} = -280 \pm 55$ mas/yr; Table \ref{tab:wise_spitzer_pm}) differs substantially from the LP 483-66 \textit{Gaia} proper motion, with a discrepancy of nearly 5$\sigma$ along the RA direction. No anomalies are evident in our WISEA 0754+0900 ch2 imaging/astrometry, so we view our \textit{WISE}+\textit{Spitzer} linear motion as superior to estimates based solely on \textit{WISE}. As a result, we consider it unlikely that WISEA 0754+0900 and LP 483-66 are associated given the available evidence.

\subsection{Notes on Individual Objects}
\label{sec:notes}

\subsubsection{WISEA J035410.03$-$572104.0}
WISEA 0354$-$5721 has $\chi^2_{motion} = 16.7$ and therefore is not motion-confirmed by our \textit{WISE}+\textit{Spitzer} astrometry analysis. Nevertheless, this source appears to have a very faint/red counterpart moving southward when comparing $z$ band images from DECaLS DR8 \citep{dey_overview} with those from Dark Energy Survey (DES) DR1 \citep{des_dr1}. The $z$ band counterpart has $z = 22.52 \pm 0.15$ (AB) according to DECaLS DR8. Assuming that this $z$ band source is indeed linked to our \textit{WISE/Spitzer} brown dwarf candidate, then the object would be most consistent with a $\sim$T7.5 phototype and a distance of $\sim$33 pc.

\subsubsection{WISEA J053535.43$-$644518.5}
\label{sec:wise_0535}
WISEA 0535$-$6445 is unusual among our sample in that its ch1$-$ch2 $\approx$ 0 color is not indicative of a mid-T or later brown dwarf, but rather is consistent with a broad range of earlier spectral types \citep{patten06}. Little archival photometry is available for this object, which lacks counterparts in \textit{Gaia} and 2MASS while falling outside of the VHS footprint. The 2MASS non-detection implies a $J - W2$ $\gtrsim$ 1.9 color limit. WISEA 0535$-$6445 has $W1-W2$ = 0.17 $\pm$ 0.04 according to the CatWISE Preliminary catalog, with slightly redder \textit{WISE} colors of $W1-W2$ = 0.23 $\pm$ 0.05 and $W1-W2$ = 0.29 $\pm$ 0.02 according to AllWISE and unWISE Catalog, respectively. The NOAO Source Catalog \citep{nsc_dr1} reports a DECam counterpart with AB magnitudes of $i$ = 21.87 $\pm$ 0.07 and $z$ = 20.36 $\pm$ 0.06. The $J - W2$ lower bound,  $W1-W2$ color and $i - z$ color indicate an early L type, and $i - W2$ = $6.61 \pm 0.08$ suggests a spectral type of $\sim$L3. The WISEA 0535$-$6445 photometry is not consistent with an M subdwarf or L subdwarf scenario. If WISEA 0535$-$6445 is an L3 dwarf, then its photometric distance estimate would be $\sim$84 pc, corresponding to a tangential velocity of $\sim$112 km/s.

\subsubsection{CWISEP J135937.65$-$435226.9}
CWISEP 1359$-$4352 was inadvertently targeted as part of p14299 despite the fact that CatWISE p14034 had previously observed it \citep{catwise_p14034}. Our p14299 observations of this object are deeper than those from p14034 and so provide a higher S/N color estimate of ch1$-$ch2 = 2.25 $\pm$ 0.09 mag, corresponding to a best-fit spectral type estimate of T9.5. The 1$\sigma$ ch1$-$ch2 color intervals from p14034 and p14299 overlap for this source, although the central color value from \cite{catwise_p14034} had scattered sufficiently redward to yield a phototype of Y0 rather than T9.5 as we have obtained here.

\section{Conclusion}
\label{sec:conclusion}

We have presented critical \textit{Spitzer} photometric follow-up of 95 brown dwarf candidates discovered by the Backyard Worlds citizen science project. Our \textit{Spitzer} astrometry allows us to verify the motions of most of these objects, certifying that these are new members of the solar neighborhood. Additionally, Keck/NIRES and Magellan/FIRE spectroscopy of ten candidates confirms in all cases that the targets are brown dwarfs. Our \textit{Spitzer} imaging also yields photometric spectral types and distances; these estimates are crucial for pinpointing the superlative objects among our sample, separating out the strong Y dwarf candidates from the late T dwarfs.

Among our most exciting discoveries are new candidate members of the 10 pc sample, two objects moving faster than 2$''$/yr, three T type subdwarf candidates, five Y dwarf candidates, and a new T8 plus white dwarf comoving system. Our Y dwarf candidates begin bridging the gap between the bulk of the Y dwarf population and the coldest known brown dwarf, making them potential targets for \textit{JWST} spectroscopy. Backyard Worlds is actively pursuing additional ground-based follow-up of the discoveries presented in this study, especially spectroscopy where feasible. 
While this work's new brown dwarf candidates already demonstrate the power of citizen science for mapping the solar neighborhood, these objects make up only a small fraction of Backyard Worlds moving object discoveries to date. As NEOWISE continues scanning the sky, Backyard Worlds will endeavor to search all of its newly delivered data for yet more cold and close neighbors to the Sun.

\begin{center}
ACKNOWLEDGMENTS
\end{center}

We thank the anonymous referee for valuable comments. The Backyard Worlds: Planet 9 team would like to thank the many Zooniverse volunteers who have participated in this project, from providing feedback during the beta review stage to classifying flipbooks to contributing to the discussions on TALK. We would also like to thank the Zooniverse web development team for their work creating and maintaining the Zooniverse platform and the Project Builder tools. We thank the Astro Data Lab team for providing catalog access to Backyard Worlds participants. This research was supported by NASA grant 2017-ADAP17-0067. SLC acknowledges the support of a STFC Ernest Rutherford Fellowship. Support for this work was provided by NASA through the NASA Hubble Fellowship grants \textit{HST}-HF2-51415.001-A and \textit{HST}-HF2-51447.001-A awarded by the Space Telescope Science Institute, which is operated by the Association of Universities for Research in Astronomy, Inc., for NASA, under contract NAS5-26555. This publication makes use of data products from the Wide-field Infrared Survey Explorer, which is a joint project of the University of California, Los Angeles, and the Jet Propulsion Laboratory/California Institute of Technology, funded by the National Aeronautics and Space Administration. This research has made use of the NASA/IPAC Infrared Science Archive, which is funded by the National Aeronautics and Space Administration and operated by the California Institute of Technology. This research has made use of the VizieR catalogue access tool, CDS, Strasbourg, France (DOI : 10.26093/cds/vizier). The original description of the VizieR service was published in \cite{vizier}. This work has made use of the white dwarf cooling models hosted by Pierre Bergeron at \url{http://www.astro.umontreal.ca/~bergeron/CoolingModels}.

This research uses services or data provided by the Astro Data Lab at NSF's National Optical-Infrared Astronomy Research Laboratory. NSF's NOIRLab is operated by the Association of Universities for Research in Astronomy (AURA), Inc. under a cooperative agreement with the National Science Foundation.

This publication uses data generated via the Zooniverse.org platform, development of which is funded by generous support, including a Global Impact Award from Google, and by a grant from the Alfred P. Sloan Foundation.

Based on observations obtained with CPAPIR at the Observatoire du Mont M\'egantic, funded by the Universit\'e de Montr\'eal, Universit\'e Laval, the Natural Sciences and Engineering Research Council of Canada (NSERC), the Fond Qu\'eb\'ecois de la Recherche sur la Nature et les Technologies (FQRNT), and the Canada Economic Development program.

The Legacy Surveys consist of three individual and complementary projects: the Dark Energy Camera Legacy Survey (DECaLS; NOAO Proposal ID 2014B-0404; PIs: David Schlegel and Arjun Dey), the Beijing-Arizona Sky Survey (BASS; NOAO Proposal ID 2015A-0801; PIs: Zhou Xu and Xiaohui Fan), and the Mayall z-band Legacy Survey (MzLS; NOAO Proposal ID 2016A-0453; PI: Arjun Dey). DECaLS, BASS and MzLS together include data obtained, respectively, at the Blanco telescope, Cerro Tololo Inter-American Observatory; the Bok telescope, Steward Observatory, University of Arizona; and the Mayall telescope, Kitt Peak National Observatory. The Legacy Surveys project is honored to be permitted to conduct astronomical research on Iolkam Du'ag (Kitt Peak), a mountain with particular significance to the Tohono O'odham Nation.

This project used data obtained with the Dark Energy Camera (DECam), which was constructed by the Dark Energy Survey (DES) collaboration. Funding for the DES Projects has been provided by the U.S. Department of Energy, the U.S. National Science Foundation, the Ministry of Science and Education of Spain, the Science and Technology Facilities Council of the United Kingdom, the Higher Education Funding Council for England, the National Center for Supercomputing Applications at the University of Illinois at Urbana-Champaign, the Kavli Institute of Cosmological Physics at the University of Chicago, Center for Cosmology and Astro-Particle Physics at the Ohio State University, the Mitchell Institute for Fundamental Physics and Astronomy at Texas A\&M University, Financiadora de Estudos e Projetos, Fundacao Carlos Chagas Filho de Amparo, Financiadora de Estudos e Projetos, Fundacao Carlos Chagas Filho de Amparo a Pesquisa do Estado do Rio de Janeiro, Conselho Nacional de Desenvolvimento Cientifico e Tecnologico and the Ministerio da Ciencia, Tecnologia e Inovacao, the Deutsche Forschungsgemeinschaft and the Collaborating Institutions in the Dark Energy Survey. The Collaborating Institutions are Argonne National Laboratory, the University of California at Santa Cruz, the University of Cambridge, Centro de Investigaciones Energeticas, Medioambientales y Tecnologicas-Madrid, the University of Chicago, University College London, the DES-Brazil Consortium, the University of Edinburgh, the Eidgenossische Technische Hochschule (ETH) Zurich, Fermi National Accelerator Laboratory, the University of Illinois at Urbana-Champaign, the Institut de Ciencies de l'Espai (IEEC/CSIC), the Institut de Fisica d'Altes Energies, Lawrence Berkeley National Laboratory, the Ludwig-Maximilians Universitat Munchen and the associated Excellence Cluster Universe, the University of Michigan, the National Optical Astronomy Observatory, the University of Nottingham, the Ohio State University, the University of Pennsylvania, the University of Portsmouth, SLAC National Accelerator Laboratory, Stanford University, the University of Sussex, and Texas A\&M University.

BASS is a key project of the Telescope Access Program (TAP), which has been funded by the National Astronomical Observatories of China, the Chinese Academy of Sciences (the Strategic Priority Research Program ``The Emergence of Cosmological Structures'' Grant XDB09000000), and the Special Fund for Astronomy from the Ministry of Finance. The BASS is also supported by the External Cooperation Program of Chinese Academy of Sciences (Grant 114A11KYSB20160057), and Chinese National Natural Science Foundation (Grant 11433005).

The Legacy Surveys imaging of the Dark Energy Spectroscopic Instrument (DESI) footprint is supported by the Director, Office of Science, Office of High Energy Physics of the U.S. Department of Energy under Contract No. DE-AC02-05CH1123, by the National Energy Research Scientific Computing Center, a DOE Office of Science User Facility under the same contract; and by the U.S. National Science Foundation, Division of Astronomical Sciences under Contract No. AST-0950945 to NOIRLab.

\vspace{5mm}

\textit{Facilities:} \textit{Spitzer}(IRAC), \textit{WISE}/NEOWISE, Keck(NIRES), Magellan(FIRE), OMM(CPAPIR), Mayall(MOSAIC-3), Blanco(DECam), UKIRT(WFCAM), VISTA(VIRCAM), Astro Data Lab, IRSA, 2MASS, \textit{Gaia}

\textit{Software:} MOPEX \citep{mopex, mopex_extraction}, WiseView \citep{wiseview}, astrometry.net \citep{astrometry_dot_net}, DESI Legacy Surveys Viewer \citep{dey_overview}, Zooniverse Project Builder

\bibliography{ms}

\begin{deluxetable}{cll}
\tablecaption{Targets Without Any \textit{Spitzer} Counterpart \label{tab:duds}}
\tablehead{
\colhead{AOR} & \colhead{designation} & \colhead{$W2$} \\
\colhead{} & \colhead{} & \colhead{(mag)}
}
\startdata
 70080512 & WISEA J014614.70$-$290048.4 & 16.49 $\pm$ 0.20 \\
 70082560 & WISEU J050511.06+304355.6 & 16.38 $\pm$ 0.12 \\
 70084096 & WISEA J132441.18$-$372024.8 & 16.25 $\pm$ 0.19 \\
 70083072 & WISEA J235136.37$-$250247.6 & 16.14 $\pm$ 0.19 \\
\enddata
\tablecomments{Each object's designation prefix corresponds to the \textit{WISE} catalog from which the quoted $W2$ magnitude has been drawn.}
\end{deluxetable}

\movetabledown=13mm
\begin{longrotatetable}
\begin{deluxetable*}{lccccccccc}
\tablecaption{Mid-infrared Photometry from \textit{WISE} and \textit{Spitzer} \label{tab:wise_spitzer_photom}}
\tablehead{
\colhead{Name} & \colhead{motion} & \colhead{$W1$} & \colhead{$W2$} & \colhead{$W1-W2$} & \colhead{AOR} & \colhead{ch1} & \colhead{ch2} & \colhead{ch1$-$ch2} & \colhead{\textit{WISE}}  \\
 & \colhead{confirmed} & \colhead{(mag)} & \colhead{(mag)}  & \colhead{(mag)} & & \colhead{(mag)} & \colhead{(mag)} & \colhead{(mag)} & \colhead{phot. ref.\tablenotemark{c}} }
\startdata
CWISE J000229.93+635217.0  & yes & $18.21 \pm 0.21$ & $15.73 \pm 0.05$\tablenotemark{b} & $2.48 \pm 0.22$ & 38724864 & $17.349 \pm 0.246$ & $15.690 \pm 0.055$ & $1.659 \pm 0.252$ & 6 \\
WISEA J001449.96+795116.1  & yes & $18.72 \pm 0.28$ & $16.00 \pm 0.06$ & $2.72 \pm 0.28$ & 65969152 & $17.752 \pm 0.059$ & $15.863 \pm 0.019$ & $1.890 \pm 0.062$ & 1 \\
WISEU J001908.31$-$094323.3  & yes & $18.33 \pm 0.17$ & $15.81 \pm 0.07$\tablenotemark{b} & $2.52 \pm 0.19$ & 70084352 & $17.119 \pm 0.034$ & $15.666 \pm 0.019$ & $1.453 \pm 0.038$ & 2 \\
WISEA J002810.59+521853.1  & yes & $18.05 \pm 0.13$ & $15.60 \pm 0.05$ & $2.45 \pm 0.14$ & 68143872 & $17.360 \pm 0.043$ & $15.669 \pm 0.019$ & $1.691 \pm 0.046$ & 1 \\
WISEA J002909.90+190516.6  & yes & $17.58 \pm 0.10$ & $15.36 \pm 0.05$ & $2.21 \pm 0.11$ & 65969664 & $16.727 \pm 0.025$ & $15.301 \pm 0.018$ & $1.426 \pm 0.032$ & 1 \\
WISEA J003631.29$-$642735.6  & no & $18.31 \pm 0.16$ & $15.86 \pm 0.05$ & $2.45 \pm 0.17$ & 70079744 & $17.203 \pm 0.037$ & $15.816 \pm 0.019$ & $1.388 \pm 0.041$ & 1 \\
WISEU J004851.21+250814.9  & yes &  -  & $15.04 \pm 0.03$\tablenotemark{b} &  -  & 68141824 & $16.793 \pm 0.026$ & $14.743 \pm 0.016$ & $2.050 \pm 0.031$ & 2 \\
WISEU J005559.88+594745.0  & yes &  -  & $15.20 \pm 0.03$\tablenotemark{b} &  -  & 68141568 & $16.450 \pm 0.021$ & $14.965 \pm 0.016$ & $1.485 \pm 0.026$ & 2 \\
\enddata
\tablenotetext{a}{The discovery of CWISEP J135937.65$-$435226.9 was presented in \cite{catwise_p14034}, but its more recent p14299 \textit{Spitzer} photometry provided here is deeper than that of \cite{catwise_p14034}.}
\tablenotetext{b}{\textit{WISE} photometry is or may be contaminated by blending.}
\tablenotetext{c}{Reference for \textit{WISE} photometry. 1 = CatWISE Preliminary catalog, 2 = unWISE Catalog, 3 = AllWISE catalog, 4 = CatWISE Preliminary catalog reject table, 5 = \cite{wise2150}, 6 = CatWISE 2020 catalog.}
\end{deluxetable*}
\end{longrotatetable}

\begin{longrotatetable}
\begin{deluxetable}{lllllllll}
\tablecaption{Details on Spectroscopic Follow-Up}
\tablehead{
\colhead{Name} &
\colhead{Telescope}&
\colhead{Instrument}&
\colhead{Date}&
\colhead{Exp(s)} &
\colhead{Frames}  &
\colhead{Telluric}  &
\colhead{Conditions} &
\colhead{SpT}\\
  \colhead{(1)}     &
  \colhead{(2)}     &
  \colhead{(3)}     &
  \colhead{(4)}     &
  \colhead{(5)}     &
  \colhead{(6)}     &
  \colhead{(7)}     &
  \colhead{(8)}     &
  \colhead{(9)}}
\startdata
WISEU J005559.88+594745.0&Keck II&NIRES& 28 Oct 2019&300&3&HD5071&Clear& T8\\
WISEA J021420.18$-$573245.1&Baade&FIRE&12 Dec 2019&120&6&HD23722&Poor&$>$T6\\
WISEA J030534.09$-$582635.5&Baade&FIRE&01 Dec 2018&120&8&HD53607&Clear&T8 \\
WISEA J050238.28+100750.0&Baade&FIRE&10 Dec 2019&120&6&HD287640&Poor&$>$T6\\
WISEA J050615.56$-$514521.3&Baade&FIRE&12 Dec 2019&120&5&HD60331&Poor&$\ge$T8\\
WISEA J053512.01$-$773829.7&Baade&FIRE&01 Dec 2018&120&8&HD90104&Clear&$\ge$T8\\
WISEA J083019.97$-$632305.4&Baade&FIRE&01 Dec 2018&120&6&HD45647&Clear&T8-T9\\
WISEU J215018.46$-$752053.0\tablenotemark{a}&Baade&FIRE&01 Dec 2018&120&6&HD200523&Clear&T8\\
WISEA J230930.58+145633.1&Keck II&NIRES&01 Sep 2018&300&12&HD7215&Clear&T8\\
WISEA J233816.47$-$732929.7&Baade&FIRE&12 Dec 2019&120&4&HD23722&Poor&$>$T6\\
\enddata 
\tablenotetext{a}{See \citet{wise2150} for details on this system.}
\label{tab:spectroscopy}
\end{deluxetable}

\end{longrotatetable}

\startlongtable
\begin{deluxetable*}{lcccccc}
\tablecaption{Recalibrated \textit{WISE} $W2$ Positions \label{tab:wise_positions}}
\tablehead{
\colhead{Name} & \colhead{RA} & \colhead{Dec} & \colhead{$\sigma_{RA}$} & \colhead{$\sigma_{Dec}$} & \colhead{time} & \colhead{mean}  \\
 & \colhead{(deg, ICRS)} & \colhead{(deg, ICRS)} & \colhead{(mas)} & \colhead{(mas)} & \colhead{slice\tablenotemark{a}} & \colhead{MJD} 
}
\startdata
CWISE J000229.93+635217.0 & 0.624757 & 63.871147 & 402 & 389 & post0\_2yr & 56983.52 \\
                           & 0.625715 & 63.871405 & 389 & 374 & post1\_2yr & 57716.24 \\
WISEA J001449.96+795116.1 & 3.708443 & 79.854488 & 432 & 402 & pre       & 55359.17 \\
                           & 3.711609 & 79.854339 & 508 & 476 & post0\_1yr & 56813.37 \\
                           & 3.712429 & 79.854352 & 639 & 590 & post1\_1yr & 57172.58 \\
                           & 3.712981 & 79.854583 & 531 & 489 & post2\_1yr & 57526.24 \\
                           & 3.713487 & 79.854231 & 519 & 483 & post3\_1yr & 57900.05 \\
                           & 3.715139 & 79.854420 & 471 & 433 & post4\_1yr & 58245.57 \\
\enddata
\tablenotetext{a}{See Table 5 of \cite{catwise_p14034} for definitions of the \textit{WISE}/NEOWISE time slices used for $W2$ astrometry.}
\end{deluxetable*}

\startlongtable
\begin{deluxetable*}{lcccccccc}
\tablecaption{\textit{Spitzer} ch2 Coordinates Used in Linear Motion Fits \label{tab:spitzer_positions}}
\tablehead{
\colhead{Name} & \colhead{AOR} & \colhead{RA} & \colhead{Dec} & \colhead{$\sigma_{RA}$} & \colhead{$\sigma_{Dec}$} & \colhead{MJD} & \colhead{method} & \colhead{$N_{calib}$} \\
 & & \colhead{(deg, ICRS)} & \colhead{(deg, ICRS)} & \colhead{(mas)} & \colhead{(mas)} &  & \colhead{number\tablenotemark{a}} & 
}
\startdata
CWISE J000229.93+635217.0  & 38724864 & 0.62231037 & 63.87123123 & 72 & 73 & 55241.26 &        1 &           46 \\
WISEA J001449.96+795116.1  & 65969152 & 3.71501259 & 79.85442342 & 40 & 37 & 58468.01 &        1 &           23 \\
WISEU J001908.31$-$094323.3  & 70084352 & 4.78470790 & $-$9.72344646 & 27 & 37 & 58797.40 &        2 &           11 \\
WISEA J002810.59+521853.1  & 68143872 & 7.04428023 & 52.31384701 & 32 & 33 & 58432.59 &        1 &           63 \\
WISEA J002909.90+190516.6  & 65969664 & 7.29110034 & 19.08729895 & 27 & 19 & 58412.28 &        2 &           12 \\
WISEA J003631.29$-$642735.6  & 70079744 & 9.13131560 & $-$64.46003825 & 20 & 26 & 58735.08 &        1 &           15 \\
WISEU J004851.21+250814.9  & 68141824 & 12.21464712 & 25.13658190 & 20 & 20 & 58422.32 &        1 &           11 \\
WISEU J005559.88+594745.0  & 68141568 & 14.00051522 & 59.79578164 & 23 & 27 & 58440.22 &        1 &          139 \\
\enddata
\tablecomments{$N_{calib}$ is the number of \textit{Gaia} DR2 sources used for \textit{Spitzer} ch2 astrometric recalibration. See  $\S$8.4.1 of \cite{catwise_p14034} for additional details.}
\tablenotetext{a}{The astrometric calibrator selection `method number' here refers to the method number listed in Table 6 of \cite{catwise_p14034}.}
\end{deluxetable*}

\movetabledown=7mm
\begin{longrotatetable}
\begin{deluxetable*}{lcccccccccccc}
\tablecaption{Linear Motion Fit Results Combining \textit{WISE} and \textit{Spitzer} Astrometry \label{tab:wise_spitzer_pm}}
\tablehead{
\colhead{Name} & \colhead{$\alpha_0$} & \colhead{$\delta_0$} & \colhead{$\sigma_{\alpha_0}$} & \colhead{$\sigma_{\delta_0}$} & \colhead{MJD$_0$} & \colhead{$\mu_{\alpha}$} & \colhead{$\mu_{\delta}$} & \colhead{$\mu_{tot}$} & \colhead{$\chi^2_{motion}$} & \colhead{$\chi^2$} & \colhead{\# dof} & \colhead{$\chi^2_{\nu}$} \\
 & \colhead{(deg, ICRS)} & \colhead{(deg, ICRS)} & \colhead{(mas)}  & \colhead{(mas)} & & \colhead{mas/yr} & \colhead{mas/yr} & \colhead{mas/yr} & & & & }
\startdata
CWISE J000229.93+635217.0  & 0.6225026 & 63.8712344 & 70 & 71 & 55379.85 & $802 \pm 49$ & $44 \pm 47$ & $803 \pm 49$ & 268.7 & 2.53 & 2 & 1.26 \\
WISEA J001449.96+795116.1  & 3.7149072 & 79.8544230 & 40 & 37 & 58416.60 & $487 \pm 44$ & $-9 \pm 41$ & $487 \pm 44$ & 123.1 & 5.15 & 10 & 0.52 \\
WISEU J001908.31$-$094323.3  & 4.7847093 & $-$9.7234432 & 28 & 37 & 58778.24 & $-126 \pm 58$ & $-286 \pm 59$ & $312 \pm 59$ & 28.5 & 4.04 & 2 & 2.02 \\
WISEA J002810.59+521853.1  & 7.0442758 & 52.3138656 & 33 & 33 & 58383.44 & $49 \pm 34$ & $-458 \pm 34$ & $461 \pm 34$ & 181.9 & 6.43 & 2 & 3.22 \\
WISEA J002909.90+190516.6  & 7.2911017 & 19.0873034 & 27 & 19 & 58391.51 & $-80 \pm 39$ & $-296 \pm 40$ & $306 \pm 40$ & 59.4 & 0.90 & 2 & 0.45 \\
WISEA J003631.29$-$642735.6  & 9.1313102 & $-$64.4600379 & 20 & 26 & 58715.36 & $143 \pm 38$ & $-22 \pm 37$ & $144 \pm 38$ & 14.6 & 0.78 & 2 & 0.39 \\
WISEU J004851.21+250814.9  & 12.2145993 & 25.1366132 & 20 & 20 & 58367.15 & $1009 \pm 21$ & $-728 \pm 21$ & $1244 \pm 21$ & 3446.9 & 8.49 & 10 & 0.85 \\
WISEU J005559.88+594745.0  & 14.0004779 & 59.7957828 & 23 & 27 & 58375.91 & $408 \pm 25$ & $-24 \pm 25$ & $409 \pm 25$ & 275.5 & 12.59 & 10 & 1.26 \\
\enddata
\tablecomments{$\chi^2_{\nu} = \chi^2/$(\# dof).}
\end{deluxetable*}
\end{longrotatetable}

\startlongtable
\begin{deluxetable*}{lcccccc}
\tablecaption{Ground-based NIR Photometry from the VISTA/UKIRT Archives and OMM Follow-up for Motion-confirmed Members of Our Sample \label{tab:jhk_phot}}
\tablehead{
\colhead{Name} & \colhead{$Y$}       & \colhead{$J_{MKO}$} & \colhead{$H_{MKO}$} & \colhead{$K_{MKO}$} & \colhead{$K_S$} & \colhead{origin} \\
               & \colhead{(mag)}     & \colhead{(mag)}     & \colhead{(mag)}     & \colhead{(mag)}     & \colhead{(mag)} &  }
\startdata
WISEA J001449.96+795116.1 &    -    & $19.72 \pm 0.62$ &    -    &    -    &    -    & CPAPIR \\
WISEU J001908.31$-$094323.3 & $19.72 \pm 0.10$ & $18.80 \pm 0.08$ &    -    &    -    & $> 18.13$ & VIRCAM \\
WISEA J002810.59+521853.1 &    -    & $19.20 \pm 0.16$ &    -    &    -    &    -    & WFCAM \\
WISEA J002909.90+190516.6 &    -    & $17.81 \pm 0.04$ &    -    &    -    &    -    & WFCAM \\
WISEU J004851.21+250814.9 &    -    & $19.51 \pm 0.16$ &    -    &    -    &    -    & WFCAM \\
WISEU J005559.88+594745.0 &    -    & $17.90 \pm 0.05$ &    -    &    -    &    -    & WFCAM \\
WISEA J010116.12$-$520629.8 &    -    & $19.76 \pm 0.12$ &    -    &    -    & $> 18.31$ & VIRCAM \\
WISEA J012057.06$-$764216.4 & $20.14 \pm 0.14$ & $19.10 \pm 0.10$ &    -    &    -    & $> 18.13$ & VIRCAM \\
\enddata
\tablecomments{Magnitude limits are $5\sigma$.}
\end{deluxetable*}

\clearpage

\begin{deluxetable}{ll}
\tablecaption{Additional OMM $J$ band Photometry \label{tab:omm_extra}}
\tablehead{
\colhead{designation} & \colhead{$J_{MKO}$} \\
\colhead{} & \colhead{(mag)}
}
\startdata
WISEA J002810.59+521853.1  & $19.58 \pm 0.58$ \\
WISEA J002909.90+190516.6  & $18.02 \pm 0.17$ \\
WISEU J005559.88+594745.0  & $16.52 \pm 0.05$\tablenotemark{a} \\
WISEA J032600.28+421056.8  & $18.85 \pm 0.28$ \\
WISEA J035733.85+070557.4  & $17.51 \pm 0.08$ \\
WISEA J040702.42+190945.8  & $17.61 \pm 0.08$ \\
WISEA J042236.95$-$044203.5  & $19.62 \pm 0.39$ \\
WISEA J050238.28+100750.0  & $> 16.84$ \\
\enddata
\tablecomments{This table reports follow-up $J$ band photometry for targets not motion confirmed and/or with higher significance $J$ band measurements from archival data sets. Magnitude limits are $5\sigma$.}
\tablenotetext{a}{The OMM photometry for WISEU 0055+5947 is contaminated by blending with a brighter nearby source, resulting in a reported magnitude that is artificially bright.}
\end{deluxetable}

\clearpage

\startlongtable
\begin{deluxetable*}{lcccccc}
\tablecaption{Derived Parameters of Motion-confirmed Objects with \textit{Spitzer} ch1 and ch2 Photometry \label{tab:derived_properties}}
\tablehead{
\colhead{Name} & \colhead{SpT} & \colhead{$M_{ch2}$} & \colhead{distance} & \colhead{$T_{\rm eff}$} & \colhead{$H_{ch2}$} & \colhead{$V_{tan}$}  \\
 & \colhead{(phototype)} & \colhead{(mag)} & \colhead{(pc)} & \colhead{(K)} & \colhead{(mag)} & \colhead{(km/s)} 
}
\startdata
CWISE J000229.93+635217.0 & 7.5 & $13.41^{+0.42}_{-0.39}$ & $28.6^{+5.6}_{-5.1}$ & $730^{+132}_{-124}$ &  $20.21^{+0.15}_{-0.14}$& $109^{+22}_{-21}$ \\
WISEA J001449.96+795116.1 & 8.0 & $13.68^{+0.31}_{-0.31}$ & $27.4^{+4.2}_{-3.7}$ & $643^{+84}_{-84}$ &  $19.30^{+0.21}_{-0.19}$& $63^{+11}_{-10}$ \\
WISEU J001908.31$-$094323.3 & 7.0 & $13.20^{+0.30}_{-0.30}$ & $31.1^{+4.6}_{-4.0}$ & $815^{+83}_{-83}$ &  $18.14^{+0.45}_{-0.37}$& $46^{+11}_{-10}$ \\
WISEA J002810.59+521853.1 & 7.5 & $13.44^{+0.30}_{-0.30}$ & $27.9^{+4.2}_{-3.7}$ & $718^{+83}_{-83}$ &  $18.99^{+0.17}_{-0.16}$& $61^{+10}_{-9}$ \\
WISEA J002909.90+190516.6 & 7.0 & $13.18^{+0.30}_{-0.30}$ & $26.6^{+4.0}_{-3.4}$ & $827^{+82}_{-82}$ &  $17.73^{+0.30}_{-0.27}$& $39^{+8}_{-7}$ \\
WISEU J004851.21+250814.9 & 8.5 & $13.89^{+0.30}_{-0.30}$ & $14.8^{+2.2}_{-1.9}$ & $588^{+82}_{-82}$ &  $20.22^{+0.04}_{-0.04}$& $87^{+13}_{-11}$ \\
WISEU J005559.88+594745.0 & 7.0 & $13.23^{+0.30}_{-0.30}$ & $22.2^{+3.3}_{-2.9}$ & $801^{+82}_{-82}$ &  $18.02^{+0.14}_{-0.13}$& $43^{+7}_{-6}$ \\
WISEA J010116.12$-$520629.8 & 8.5 & $13.82^{+0.31}_{-0.31}$ & $23.7^{+3.7}_{-3.2}$ & $607^{+84}_{-83}$ &  $19.96^{+0.14}_{-0.13}$& $80^{+13}_{-12}$ \\
\enddata
\tablecomments{Photometric spectral type values are defined such that SpT = 6 for T6, SpT = 7 for T7,... SpT = 11 for Y1. WISEA J053535.43$-$644518.5 is omitted from this table despite being motion-confirmed, since its relatively blue ch1$-$ch2 $\approx$ 0 color does not provide a well-constrained \textit{Spitzer}-based phototype. WISEU J224547.11$-$433341.5 has been omitted because it lacks a ch1 magnitude and hence a \textit{Spitzer} phototype.}
\tablenotetext{a}{\cite{w0830} presents a trigonometric parallax for WISEA J083011.95+283716.0 yielding more reliable estimates for its phototype, $M_{ch2}$, distance, $T_{\rm eff}$, and $V_{tan}$.}
\end{deluxetable*}

 \ 

\end{document}